\documentclass[traditabstract,a4paper]{aa}
\usepackage{graphicx,txfonts,latexsym}
\usepackage{natbib,twoopt}
\usepackage{subfigure}
\usepackage{hyphenat}
\usepackage[latin1]{inputenc}
\bibpunct{(}{)}{;}{a}{}{,}

\begin{document}
\title{Domain of validity for pseudo-elliptical NFW lens models.}
\subtitle{Mass distribution, mapping to elliptical models, and arc cross section}

\author{H.~S.~D\'umet-Montoya
\and G.~B.~Caminha
\and M. ~Makler}
\institute{Instituto de Cosmologia, Relatividade e Astrof\'isica --- ICRA\\
Centro Brasileiro de Pesquisas F\'isicas\\
Rua Dr. Xavier Sigaud 150, CEP 22290-180, Rio de Janeiro, RJ, Brazil\\
and\\
Laborat\'orio Interinstitucional de e-Astronomia --- LIneA, \\
Rua Gal. Jos\'e Cristino 77, CEP 20921-400, Rio de Janeiro, RJ, Brazil}

\abstract
{\emph{Context.} Owing to their computational simplicity, models with elliptical potentials (pseudo-elliptical) are often used in gravitational lensing applications, in particular for mass modeling using arcs and for arc statistics.  However, these models generally lead to negative mass distributions in some regions and to dumbbell-shaped surface density contours for high ellipticities.

\emph{Aims.} We revisit the physical limitations of the pseudo-elliptical Navarro--Frenk--White (PNFW) model, focusing on the behavior of the mass distribution close to the tangential critical curve, where tangential arcs are expected to be formed.  We investigate the shape of the mass distribution on this region and the presence of negative convergence. We obtain a mapping from the PNFW to the NFW model with elliptical mass distribution (ENFW). We compare the arc cross section for both models, aiming to determine a domain of validity for the PNFW model in terms of its mass distribution and for the cross section.

\emph{Method.} We defined a figure of merit to {\it i}) measure the deviation of the iso-convergence contours 
of the PNFW model to an elliptical shape, {\it ii}) assigned an ellipticity $\varepsilon_\Sigma$ to these contours, {\it iii}) defined a corresponding iso-convergence contour for the ENFW model. We computed the arc cross section using the ``infinitesimal circular source approximation''.

\emph{Results.} We extend previous work by investigating the shape of the mass distribution of the PNFW model for a broad range of the potential  ellipticity parameter $\varepsilon$ and characteristic convergence $\kappa_s^\varphi$.
We show that the maximum value of  $\varepsilon$ to avoid dumbbell-shaped mass distributions is explicitly dependent on $\kappa_s^\varphi$, with higher ellipticities  ($\varepsilon \simeq 0.5$, i.e., $\varepsilon_\Sigma \simeq 0.65$) allowed for small  $\kappa_s^\varphi$. We determine a relation between the ellipticity of the mass distribution $\varepsilon_\Sigma$ and $\varepsilon$ valid for any ellipticity. We also derive the relation of characteristic convergences, obtaining a complete mapping from PNFW to ENFW models, and provide fitting formulae for connecting the parameters of both models.
Using this mapping, the cross sections for both models are compared, setting additional constraints on the parameter space of the PNFW model such that it reproduces the ENFW results.  We also find that the negative convergence regions occur far from the arc formation region and should therefore not be a problem for studies with gravitational arcs.

\emph{Conclusions.} We conclude that the PNFW model is well-suited to model an elliptical mass distribution on a larger
$\varepsilon$--$\kappa_s^\varphi$  parameter space than previously expected.   However, if we require the PNFW model to reproduce the arc cross section of the ENFW well, the ellipticity is more restricted, particularly for low $\kappa_s^\varphi$. The determination of a domain of validity for the PNFW model and the mapping to ENFW models could have implications for the use of PNFW models for the inverse modeling of lenses and for fast arc simulations, for example.}

\keywords{Gravitational lensing: strong -- Galaxies: clusters: general  -- Galaxies: halos -- dark matter} 
\titlerunning{Domain of validity for pseudo-elliptical NFW lens models}
\authorrunning{H.~S.~D\'umet-Montoya, G.~Caminha, \and M.~Makler}
\maketitle

\section{Introduction \label{introd}}

Gravitational arcs are powerful probes of  the mass distribution in galaxies \citep{Koopmans09,Barnabe11,Suyu12} and galaxy clusters \citep{kovner89,miralda93,hattori97} and can be used to constrain cosmological models \citep{bartelmann1998,oguri01, golse2002, bartelmann03,2010Sci...329..924J}.
The main techniques employed to extract information from gravitational arcs have been {\it arc statistics} \citep[i.e. counting the number of arcs in  lens samples,][]{ wu93,1994ApJ...431...74G,bartelmann94} and {\it inverse modeling} \citep[i.e. ``deprojecting'' the arcs in individual lens systems to determine the lens and the source,][]{1993A&A...273..367K,gravlens,golse2002,lensview,jullo07,2010Sci...329..924J}. The first requires large samples of arcs while the second needs detailed information on the lensing systems, typically imaging from space and lens and source redshifts.

These applications have triggered arc searches in surveys covering large areas \citep{rscs, sdss1,legacy,CS82,more11,CASSOWARYmethod, SBAS7},  in large spectrocospic surveys \citep{SLACSV,BELLSI}, and in surveys targeting known clusters \citep{luppino, lcdcs,2005MNRAS.359..417S,sdss2, sogras10,2010MNRAS.404..325R,kausch2010, sogras}. Upcoming wide field imaging surveys, such as the Dark Energy Survey \citep{des05,annis05} and the Large Synoptic Survey Telescope \citep{lsst,lsst1}, will 
lead to the identification of larger samples of arcs in thousands of galaxies and galaxy clusters, well suited for arc statistics. Moreover, deep observations from space combined with massive spectroscopy have been obtained for a limited number of clusters and were used for detailed mass modeling \citep[see, e.g., ][]{2005ApJ...621...53B,2010MNRAS.402L..44R,2010Sci...329..924J}.

The simplest models that can account for some observed properties of arcs (multiplicity, relative positions, morphology) are built from axial models by introducing an ellipticity either on the mass distribution  \citep[elliptical models,][]{schramm90,barkana98,gravlens,oguri03} or on the lensing potential \citep[pseudo-ellitpical models,][]{kochanek89,kassiola93,kneib01}. Most parametric analyses of arcs, both for the inverse modeling \citep{jullo07} and for arc statistics \citep{oguri02, oguri03}, involve one or more elliptical/pseudo-elliptical models (adding, in some cases, external shear and substructures). 

Elliptical models, whose surface density is constant over ellipses, are more realistic than pseudo-elliptical ones. For example, elliptical models are motivated by the results of N-body simulations, which show that dark matter halos are triaxial \citep{2002ApJ...574..538J,2007MNRAS.378...55M}, such that their overall mass distribution can be modeled at first order by ellipsoids, whose surface density contours are elliptical. 
In contrast, the surface density of pseudo-elliptical models  generally has a pathological behavior, exhibiting regions where it takes negative values \citep{kochanek89} and presenting a ``dumbbell'' (or ``peanut'') shape for high ellipticities \citep{kovner89, SEF,kassiola93}, which does not represent the mass distribution of most physical systems.

On the other hand, pseudo-elliptical models provide simple analytic solutions for some lensing quantities,  allowing for fast numerical methods to be implemented, whereas elliptical models require the computation of integrals, which are more demanding numerically \citep{schramm90,gravlens}. Therefore, studies that require numerous evaluations of the lensing quantities often employ pseudo-elliptical models. For example, several studies using arc simulations have used these models \citep{oguri02, mene03,mene07}. Popular codes for lens inversion \citep{golse2002,lensview, jullo07} are implemented using this type of model, too.

It is therefore relevant to determine a ``domain of validity'' for pseudo-elliptical models such that 
 the negative convergence appears far from the arc-forming region and the shape of their mass distribution is closer to elliptical. The determination of  such a validity region could be useful, for example, to evaluate if a set of model parameters derived from the inversion of a system with arcs  using the PNFW model is physically acceptable.  Even for methods that use the ENFW for lens inversion \citep{gravlens,Suyu12}, the PNFW could be useful, within its domain of validity, for a faster coarser probing of the parameter space that would subsequently need to be refined with the elliptical model.

Once a pseudo-elliptical model is found to be acceptable, it is nevertheless necessary to provide a correspondence to  an elliptical model. This would be necessary, for example, to compare results derived from the mass modeling using observed arcs with theoretical predictions. We therefore need to establish a mapping among the parameters of the two models. This mapping could also be used to replace an elliptic model by its corresponding pseudo-elliptic in lensing simulations.

In this paper we  focus on the widely used Pseudo-Elliptical Navarro--Frenk--White (hereafter PNFW) model and investigate its domain of validity as well as the mapping to  the corresponding elliptical model  (hereafter ENFW). To test the equivalence among the two models we compute the cross section for arc formation, from which we obtain additional constraints on the model parameters.

The outline of this paper is as follows: In Sect.~\ref{PNFW} we briefly review the PNFW lens model, introducing the conventions and parameter ranges to be used throughout this work.
In Sect.~\ref{phys_lim_pnfw} we discuss the region of arc formation and derive physical limits of the PNFW mass distribution in this region.  In Sect.~\ref{mapping} we consider two ways for assigning an ellipticity to the PNFW surface density. In Sect.~\ref{charac_convg_relation} we obtain a mapping from the PNFW to the ENFW models. In Sect.~\ref{compar_sc} we compare the arc cross sections of the two models using the mapping relations. In Sect.~\ref{s&c} we present the summary and concluding remarks. In Appendix~\ref{ap1} we present useful relations to derive some lensing functions for pseudo-elliptical models. In Appendix~\ref{useful-fit} we provide fitting formulae for the limits on the mass distribution of the PNFW model and for the mapping to the ENFW.

\section{\label{PNFW}The pseudo--elliptical NFW lens model}

\subsection{Circular NFW lens model}

From N-body simulations Navarro, Frenk \& White (NFW) found that the
radial (i.e. angle-averaged) density profile of dark matter haloes approximately follows the universal function \citep{nfw96,nfw97}

\begin{equation}
 \rho(r)=\frac{\rho_s}{(r/r_s)(1+r/r_s)^2},
 \label{perfil_nfw}
\end{equation}
where  $r_s$ is the scale radius and $\rho_s$ is the characteristic density of the halo. This profile has been widely used to represent the  dark matter mass distribution galaxy  to cluster scales and will be employed throughout this work.

From the density profile (\ref{perfil_nfw}), defining the dimensionless Cartesian coordinates in the lens plane%
\footnote{In Eq.~(\ref{perfil_nfw}) $r=\sqrt{\xi^2+z^2}$, where $z$ is the coordinate on the lens--observer direction and $\xi =|\vec{\xi}|$ is the coordinate perpendicular to the line-of-sight (lens plane).},
 $\vec{x}=\vec{\xi}/r_s$, the convergence, deflection angle, shear, and lensing potential are given by \citep{bartelmann96}:
\begin{eqnarray}
\kappa(x) &=& 2\, \kappa_s\,F(x) \label{kappa_nfw},\\
\alpha(x) &=& 4\,\kappa_s\frac{g(x)}{x}\label{alpha_nfw},\\
\gamma(x) &=& 2\,\kappa_s\left(\frac{2\,g(x)}{x^2}-F(x)\right) \label{shear_nfw},\\
\varphi(x)&=& 2\,\kappa_s h(x),\label{pot_nfw}
\end{eqnarray}
\noindent where the functions $F(x)$, $g(x)$  and $h(x)$ are defined in \citet[][hereafter GK02]{gk03} and the characteristic convergence, $\kappa_s$, is given by
\begin{equation}
\kappa_s=\frac{\rho_s\,r_s}{\Sigma_{\mbox{\tiny crit}}},
\label{ks_nfw}
\end{equation}
\noindent with the critical
surface mass density 
\begin{equation}
 \Sigma_{\rm crit}=\frac{c^2}{4\pi\,G}\frac{D_{OL}}{D_{LS}\,D_{OS}},
\label{Sigmacrit}
\end{equation}
where $D_{OL}$, $D_{LS}$ and $D_{OS}$ are the angular-diameter distances between the observer and lens (at redshift $z_L$), lens and source (at redshift $z_S$), and observer and source, respectively \citep{SEF,morro}.

Notice that with the choice of dimensionless coordinates $\vec{x}$, the lensing functions are independent of $r_s$. Naturally, dimensional quantities have to be scaled accordingly to recover their physical units. For example, angular quantities have to be multiplied by $r_s/D_{OL}$ to be in radians.

\subsection{Pseudo-elliptical models} \label{pemodels}
To construct ``elliptical models" from a given radial density profile, the radial coordinate $x$ is replaced by
\begin{equation}
 x_\varepsilon =
\sqrt{a_{1}\,x^2_1+a_{2}\,x^2_2},
\label{subti-ellip}
\end{equation}
such that the ellipticity is given by 
\begin{equation}
\varepsilon_\varphi = 1-\sqrt{\frac{a_{1}}{a_{2}}},
\label{ellip-pot}
\end{equation}
where we assume that $a_2 > a_1$, such that the major axis of the ellipse is along $x_1$.

This substitution can be performed on the projected mass distribution, $\kappa(x)\rightarrow \kappa(x_\varepsilon)$, leading to elliptical mass distributions \citep{bourassa73,bourassa75,bray84}. 
An alternative is to construct pseudo-elliptical models, by making the substitution in the lensing potential, $\varphi(x)\rightarrow \varphi(x_\varepsilon)$, which leads to simple analytic expressions for the lensing functions
\citep[GK02,][]{kassiola93,oguri02,jullo07}. In particular the convergence and the components of the shear
are given by (see Appendix~\ref{pe_functions})
\begin{eqnarray}
\kappa_\varepsilon(\vec{x})&=&\mathcal{A}\,\kappa(x_\varepsilon)
-\mathcal{B}\,\gamma(x_\varepsilon)\cos{2\phi_\varepsilon}
\label{kappa_pnfw},\\
\gamma_{1\varepsilon}(\vec{x})& = &
\mathcal{B}\,\kappa(x_\varepsilon)-\mathcal{A}\,
\gamma(x_\varepsilon)\cos{2\phi_\varepsilon}, \\
\gamma_{2\varepsilon}(\vec{x})& =
&-\sqrt{\mathcal{A}^2-\mathcal{B}^2}\,
\gamma(x_\varepsilon)\sin{ 2\phi_\varepsilon},\\
\gamma_\varepsilon^2(\vec{x}) & = &
\mathcal{A}^2\gamma^2(x_\varepsilon)-2\mathcal{A}\,
\mathcal {B}\,
\kappa(x_\varepsilon)\gamma(x_\varepsilon)\cos{2\phi_\varepsilon}
\nonumber  \\ &  &
+\mathcal{B}^2[\kappa^2(x_\varepsilon)-\sin^2{2\phi_\varepsilon}
\gamma^2(x_\varepsilon) ]\label{gamma_pnfw},
\end{eqnarray}
where $\mathcal{A}=\frac{1}{2}(a_{1}+a_{2})$,
$\mathcal{B}=\frac{1}{2}(a_{1}-a_{2})$, and $\kappa(x_\varepsilon)$
and $\gamma(x_\varepsilon)$ are the convergence and shear of a circular model evaluated at $x_\varepsilon$, respectively.

The PNFW model is obtained by introducing the ellipticity on the NFW lens potential (Eq.~\ref{pot_nfw}), where we denote the characteristic convergence (Eq.~\ref{ks_nfw}) by $\kappa_s^\varphi$. 

In this paper we adopt the convention \citep[][GK02]{kochanek89}
\begin{equation}
 a_{1}=1-\varepsilon, \quad a_{2}=1+\varepsilon,
\label{choice_gk02}
\end{equation}
such that the ellipticity of the lensing potential (Eq.~\ref{ellip-pot}) is given by
\begin{equation}
\varepsilon_\varphi = 1-\sqrt{\frac{1-\varepsilon}{1+\varepsilon}},
\label{ephi}
\end{equation}
and the ellipticity parameter $\varepsilon$ is defined in the range $ 0 \leq \varepsilon < 1$.

Using this convention, Eqs.~(\ref{kappa_pnfw}) and (\ref{gamma_pnfw}) yield Eqs.~(17) and (19) of GK02\footnote{After accounting for the known typo, in Eq.~ (19) of GK02 ($\cos^2{2\phi_\varepsilon} \rightarrow \sin^2{2\phi_\varepsilon}$). We thank the authors for pointing this out on the {\it lenstool} code \citep{jullo07} at {\texttt http://www.oamp.fr/cosmology/lenstool/}.}.
Another common choice of parameterization is $a_{1}=1/(1-\varepsilon)$, $a_{2}=1-\varepsilon $.
In this case the ellipticity of the potential is simply given by $\varepsilon_\varphi=\varepsilon$ and Eqs.~(\ref{kappa_pnfw}-\ref{gamma_pnfw}) yield the same expressions for the lensing functions as, for example, in \citet{lima_2009}\footnote{After rotating the lens by $\pi/2$ to follow their convention.}.

To obtain the  maximum value of $\kappa_s^\varphi$ to be used in this paper we considered the most extreme lensing systems expected in nature, with\footnote{Where $M_{200}$ is defined as the mass contained within a radius $r_{200}$ enclosing a region with mean density $200$ times the critical density of the Universe  at $z_L$ and $h$ is the Hubble parameter in units of 100 Mpc km$^{-1}$s.}
 $M_{200} = 4 \times 10^{15} h^{-1} M_{\sun}$, $z_L=1.6$, $z_S=7$ \citep{oguri2009}. 
The values of $r_s$ were obtained from the distribution of the concentration parameter $ c=r_{200}/r_{s}$, $p(c|M,z=0)$, derived from N-body simulations by \citet{angelo_paper} with redshift scaling given in \citet{2008MNRAS.391.1940M}. From  $M_{200}$, $c$, $z_L$, and $z_S$ we obtained the characteristic convergence by applying the relations for the spherical NFW model \citep[Eqs.~\ref{perfil_nfw}, \ref{pot_nfw}, and \ref{Sigmacrit}, see, e.g.,][]{caminha09}, where we chose the $\Lambda$CDM matter and cosmological constant  density parameters as $\Omega_m=0.3$ and $\Omega_\Lambda= 0.7$, respectively.
We generated many realizations of the $c$--$M_{200}$ relation, converted them to characteristic convergence, and took the 95\% upper limit of the derived $\kappa_s^\varphi$ distribution as its maximum value, which leads to $\kappa_s^\varphi \simeq 1.5$. This sets an inclusive upper limit for $\kappa_s^\varphi$, at least within the  $\Lambda$CDM framework. 
\begin{figure*}[!ht]
\centering \sidecaption \resizebox{\hsize}{!}{
\subfigure{\includegraphics{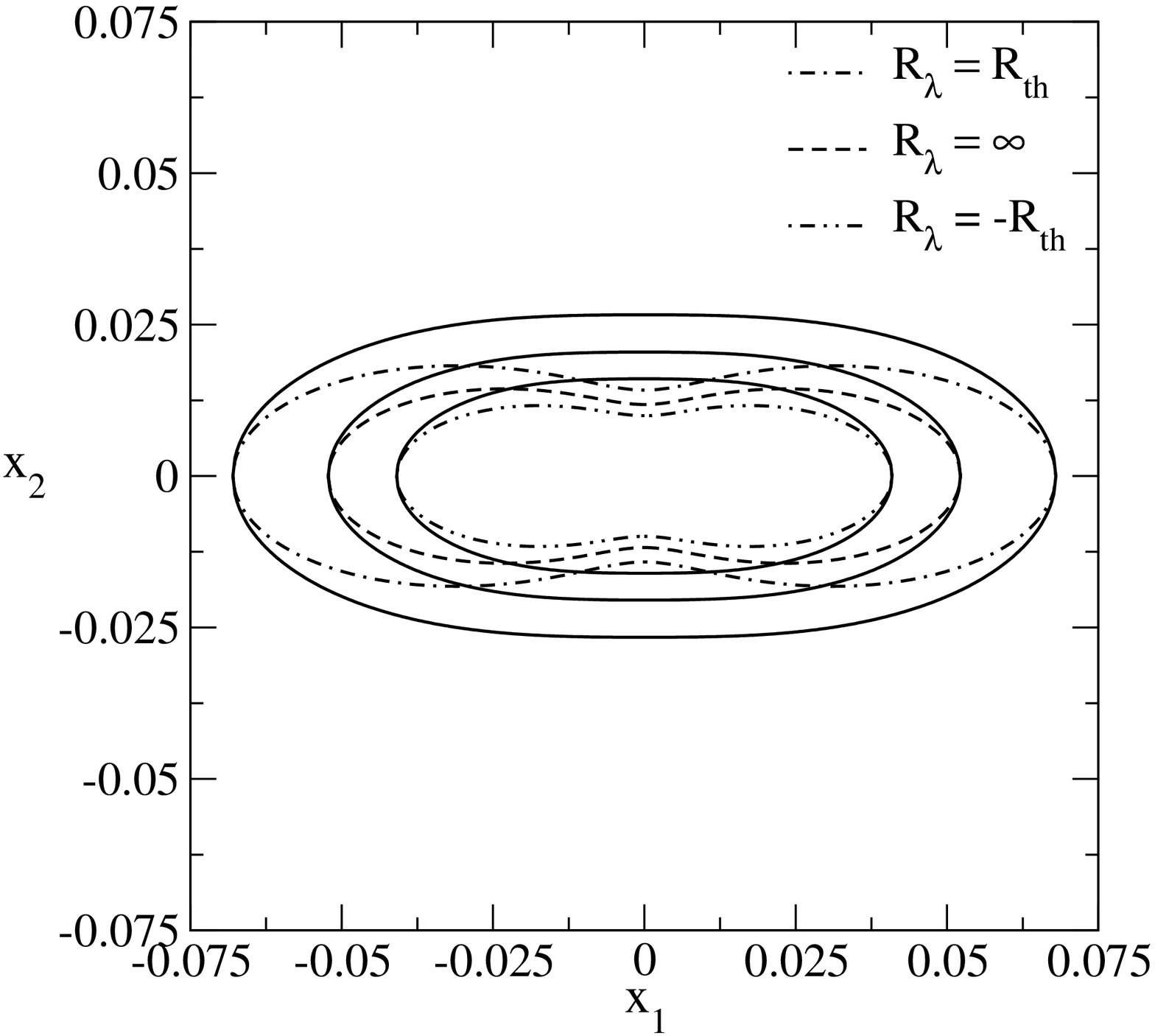}} \hspace{1.cm}
\subfigure{\includegraphics{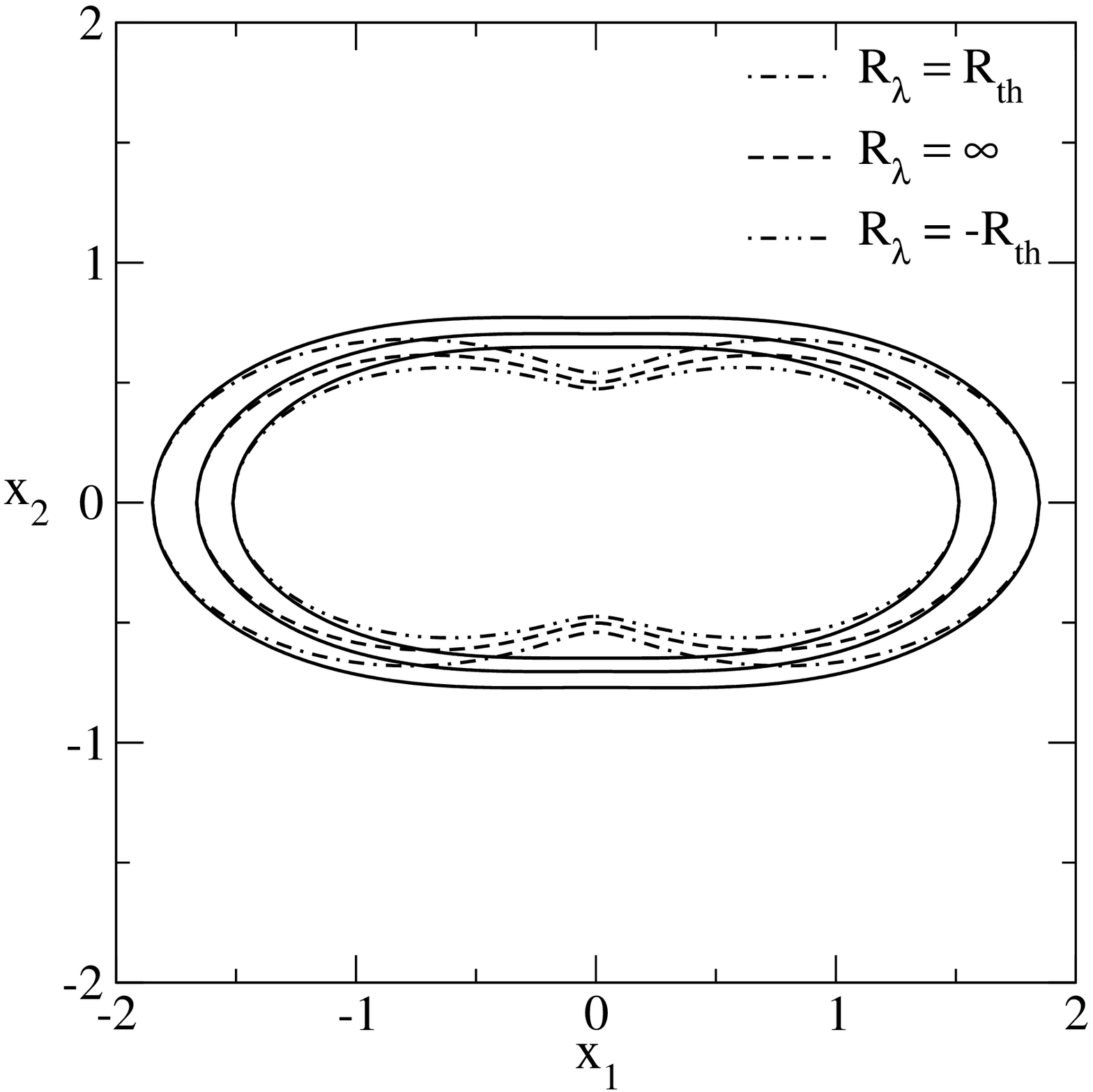}} \hspace{1.cm}
\subfigure{\includegraphics{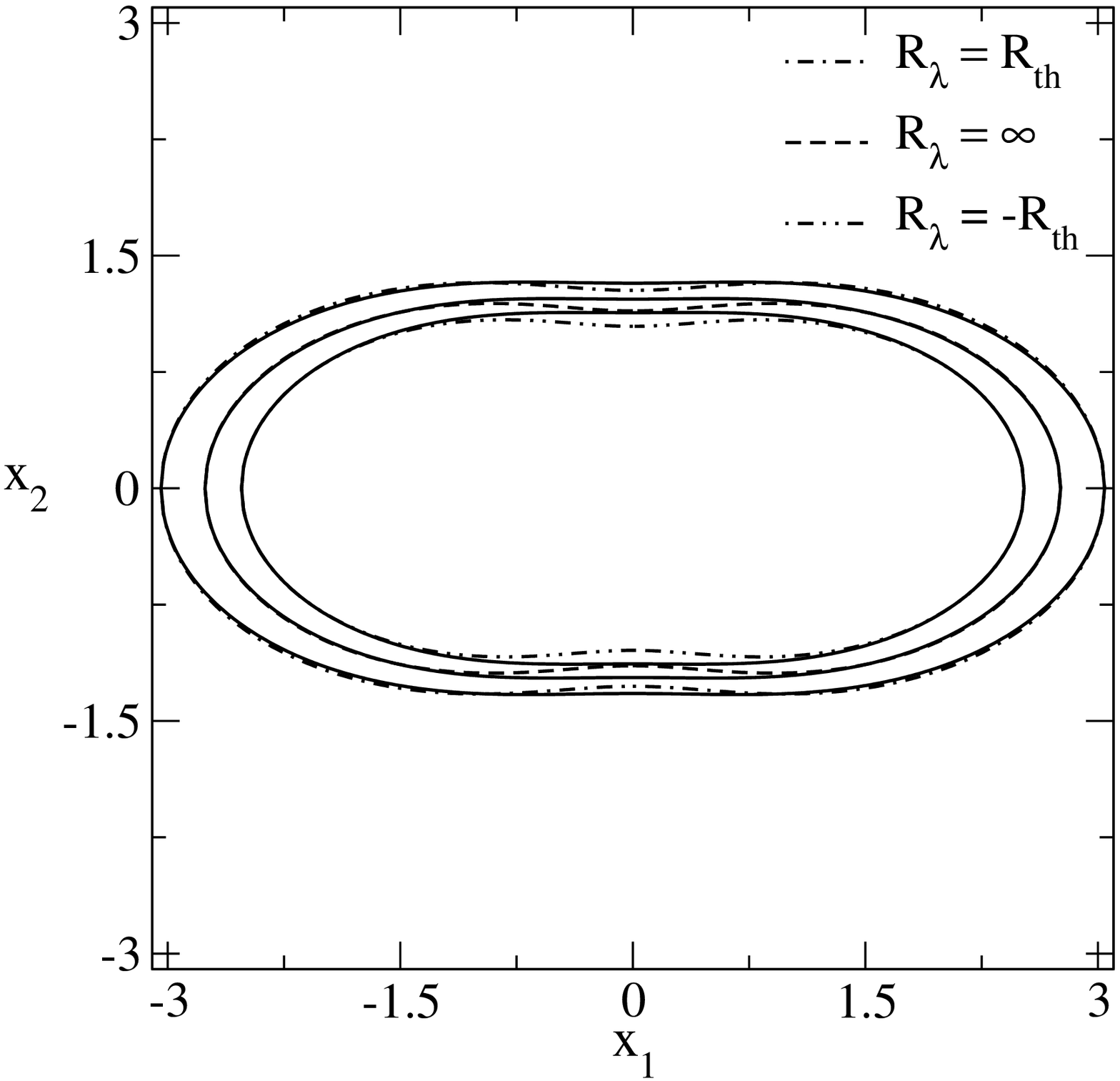}}}\caption{Critical curve ($R_\lambda =\infty$), curves of constant distortion ($R_\lambda=\pm R_{\rm th}$) and $\kappa_\varepsilon$ contours associated to each $R_\lambda$ curve (solid lines) for $R_{\rm th}=10$ and 
$\kappa_s^\varphi = 0.1$ and $\varepsilon =0.45$ (left panel), $\kappa_s^\varphi = 0.8$ and $\varepsilon =0.35$ (middle panel) and $\kappa_s^\varphi = 1.5$ and $\varepsilon =0.3$ (right panel). The axes are in units of $r_s$.} \label{nfw_curves}
\end{figure*}

\section{Physical limits of the PNFW mass distribution\label{phys_lim_pnfw}}

We focused on the mass distribution in the vicinity of the region of tangential arc\footnote{Radial arcs are more difficult to observe because they are hidden by the light of the lens, since they are formed in the central region of the lenses and are usually fainter images \citep{miralda91,bartelmann02}.} formation. The arcs are usually defined as an image 
with length-to-width ratio $L/W$ above a given threshold $R_{\rm th}$. For infinitesimal circular sources this ratio can be determined from the radial and tangential eigenvalues of the Jacobian matrix of the lens mapping, $\lambda_r$  and $\lambda_t$, respectively
\citep{wu93,bartelmann94,haman97}: 
\begin{equation}
\frac{L}{W} =\left|  R_\lambda \right |, \label{lw}
\end{equation}
where $R_\lambda := \lambda_r/\lambda_t$, with $\lambda_r=1-\kappa+\gamma$  and $\lambda_t=1-\kappa-\gamma$. 

Fixing a value for $R_{\rm th}$ determines a region limited  by the curves $R_\lambda=\pm R_{\rm th}$ (constant distortion curves), where gravitational arcs are expected to be formed. 
Although condition (\ref{lw}) does not hold for images of sources crossing the tangential caustic \citep[merger arcs,][]{rozo08,cunha10}, nor for large or noncircular sources, the curves defined above still provide a typical  scale for the region of arc formation. A common choice for the threshold is $R_{\rm th}=10$, which we adopted in this work (unless explicitly stated otherwise). The tangential critical curve is given by the condition $R_\lambda=\infty$. In Fig.~\ref{nfw_curves} these curves are shown for a few combinations of $\kappa_s^\varphi$ and $\varepsilon$. 

Once a domain for arc formation has been established, we need to associate to this region iso-convergence contours, denoted by $\kappa_\varepsilon$ contours, which define the shape of the mass distribution. We chose to match the $\kappa_\varepsilon$ contours to the $R_\lambda=\pm R_{\rm th}$ curves at the major axis ($x_2 =0$), since most arcs are expected to be formed close to this region \citep[see, e.g.,][]{dalal04,comerford06,mene07, more11}. This definition is illustrated in Fig.~\ref{nfw_curves}, where distortion curves and their associated iso-convergence contours are shown. We will use this choice throughout this paper and refer to the $\kappa_\varepsilon$ contour  by the value of  $R_\lambda$ associated to it.
\begin{figure*}[!ht]
\centering \sidecaption \resizebox{\hsize}{!}{
\subfigure{\includegraphics{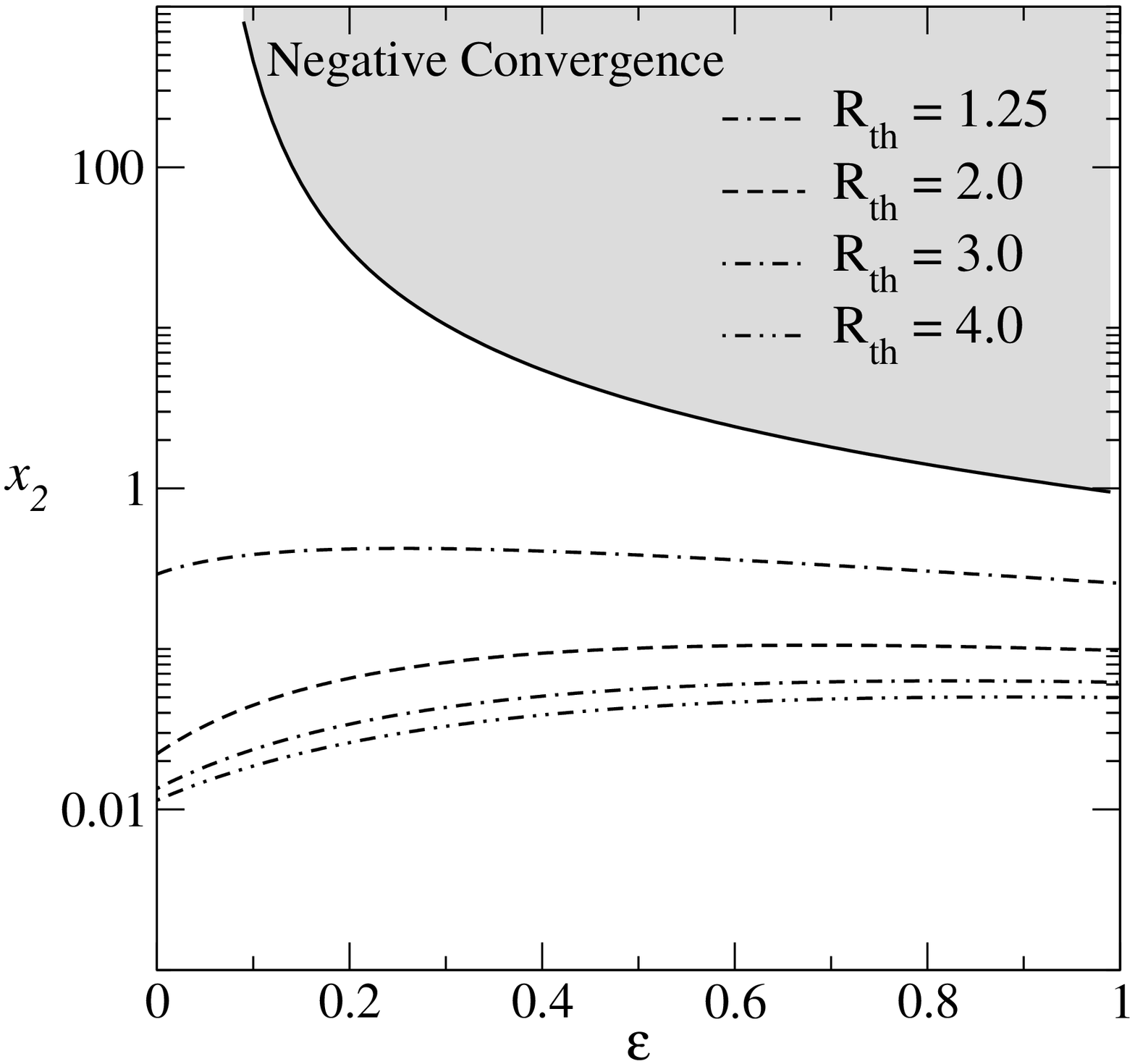}} \hspace{1.5cm}
\subfigure{\includegraphics{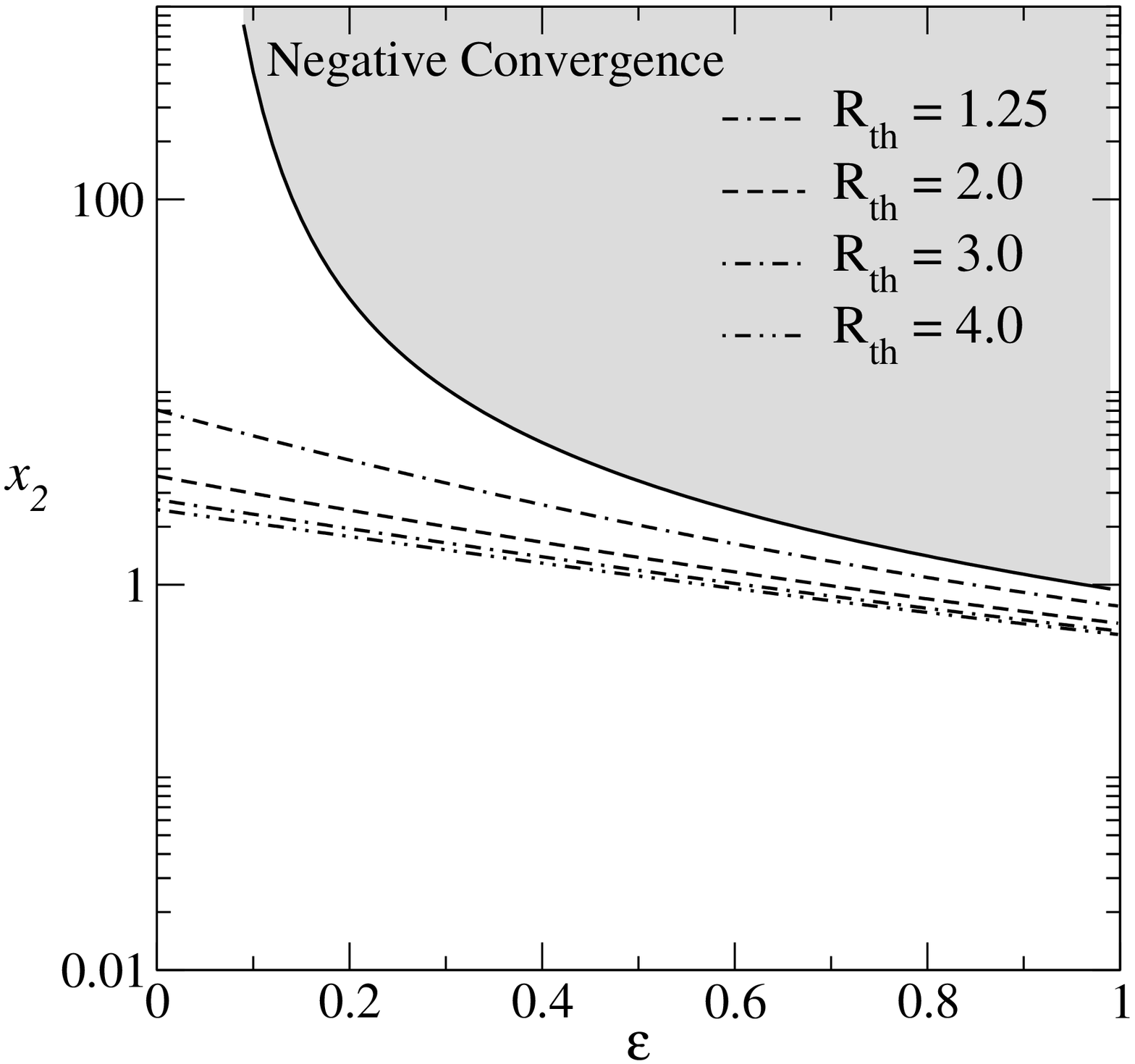}}}
\caption{\label{kappa_neg}  Negative convergence of the PNFW model. Solid lines show the intersection of $\kappa_\varepsilon=0$ contours with $x_2$ axis. Other lines correspond to the intersection of the iso-convergence contours associated to $R_\lambda=R_{\rm th}$ curves with the $x_2$ axis. Left panel: $\kappa_s=0.1$. Right panel: $\kappa_s=1.5$.}
\end{figure*}

 A fundamental problem of pseudo-elliptical models is the presence of regions with negative mass distribution \citep{kassiola93}.
For the PNFW model parameterized as in Eq.~(\ref{choice_gk02}), negative $\kappa_\varepsilon$ contours form lobes oriented along the $x_2$ axis. These regions occur far from the lens center and for any $\varepsilon>0$.
The $\kappa_\varepsilon=0$ contours are independent of  $\kappa_s^\varphi$ and the nearest point to the region of tangential arc formation is located on the $x_2$ axis.  For example, in Fig. \ref{nfw_curves},  for $\varepsilon = 0.3,0.35$, and $0.45$, negative values of $\kappa_\varepsilon$ arise at $x_2 = \pm 10.5, 7.3$ and $4.3$, respectively,  well outside the range of these plots.

We have verified that the $\kappa_\varepsilon=0$ contours do not intersect the iso-convergence contours associated to the $|R_\lambda|=R_{\rm th}$ curves in the whole $\varepsilon$ and $\kappa_s^\varphi$ range, as can be seen on Fig.~\ref{kappa_neg}. As expected, the negative convergence lobes approach the $R_{\lambda}= const.$ curves as the ellipticity increases, but never intersects these regions, even down to $R_{\rm th}=1.25$. The lower the characteristic convergence, the farther is the negative convergence region from the arc formation region. Therefore, the formation of tangential arcs occurs far from the $\kappa_\varepsilon<0$ regions, and can be ignored in the context of this paper.

\begin{figure}[!ht]
\centering \sidecaption \resizebox{\hsize}{!}{
\subfigure{\includegraphics{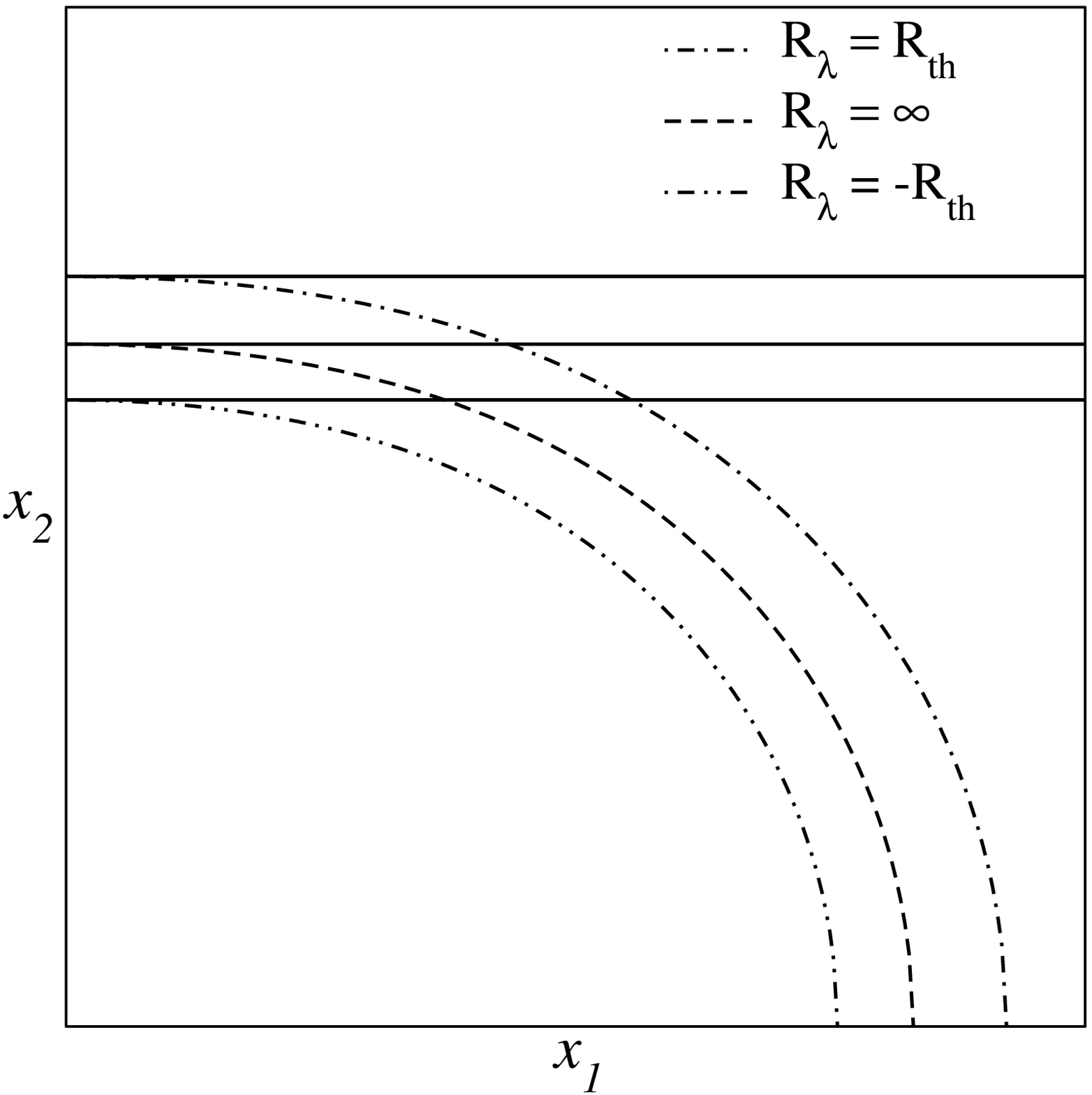}} \hspace{1.5cm}
\subfigure{\includegraphics{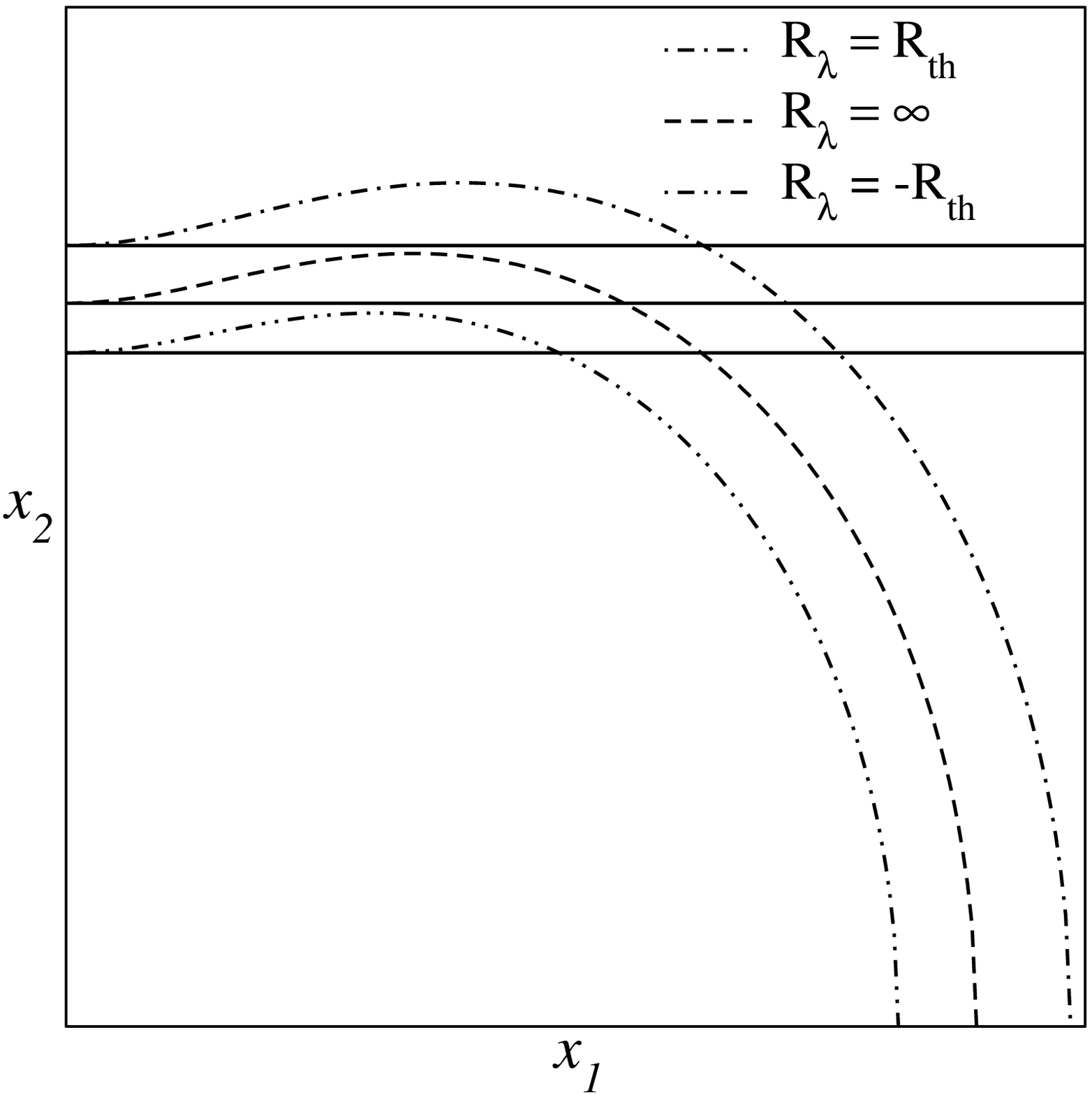}}}
\caption{\label{schemax} Sketch of the method for determining the maximum value of $\varepsilon$ to avoid dumbell-shaped $\kappa_\varepsilon$ contours associated to the $R_\lambda = const.$ curves. Left panel: Shape of the iso-convergence contours for $\varepsilon<{\varepsilon}_{\rm max}$. Right panel: Shape of the iso-convergence contours for $\varepsilon>{\varepsilon}_{\rm max}$.}
\end{figure}

Now we turn to the shape of the  $\kappa_\varepsilon$ contours. Given the chosen orientation of the major axis, for contours that are close to elliptical, the maximum value of $x_2$ ($x_2^{\rm max}$) is located at $x_1=0$. On the other hand, for dumbbell-shaped contours $x_2^{\rm max}$ is located at $x_1\neq 0$.  This simple property can be used to determine the maximum value of $\varepsilon$  ($\varepsilon_{\rm max}$) such that for $\varepsilon > \varepsilon_{\rm max}$ a dumbbell-shape emerges.
This is shown schematically in Fig.~\ref{schemax}. For a given $\kappa_s^\varphi$, we started with a low value of $\varepsilon$ and computed $x_2^{\rm max}$ for the $\kappa_\varepsilon$ contours associated to each $R_\lambda$ curve. As the ellipticity parameter is increased, $\varepsilon_{\rm max}$ is attained when the point corresponding to $x_2^{\rm max}$ starts to be located at $x_1\neq 0$. Repeating this procedure for any value of $\kappa_s^\varphi$ and $R_\lambda$ leads to the function $\varepsilon_{\rm max}(\kappa_s^\varphi,R_\lambda)$.

\begin{figure}[!htb]
\begin{center}
\sidecaption \resizebox{\hsize}{!}{\includegraphics{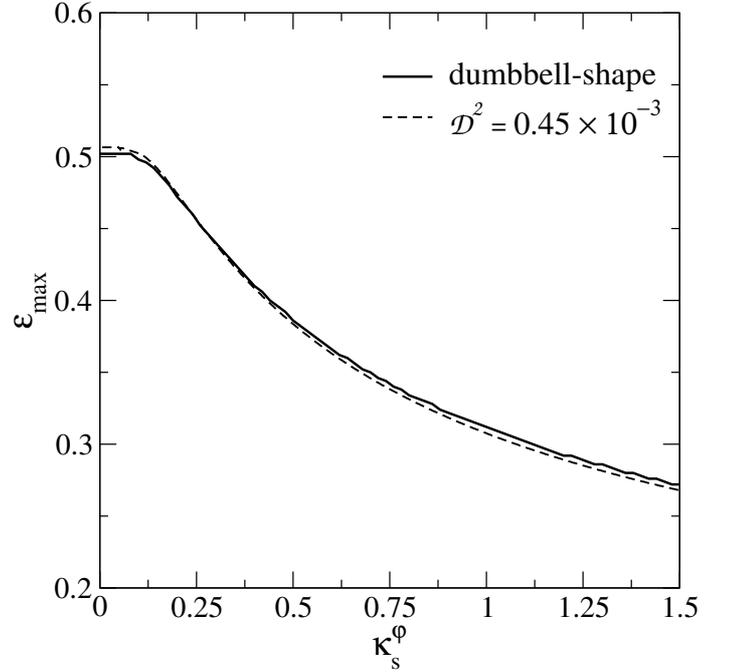}}
\caption{ \label{gof2} Maximum value of the ellipticity to avoid dumbbell-shaped mass distributions in the region close to the $R_\lambda = \infty$ contour  as a function of $\kappa_s^\varphi$. The solid line corresponds to $\varepsilon_{\rm max}$ obtained  from the procedure in Sect.~\ref{phys_lim_pnfw}. The dashed line corresponds to the values of $\varepsilon$ obtained from upper limits of the figure-of-merit ${\mathcal{D}}^2$ representing the fractional deviation of the $\kappa_\varepsilon$ contours with respect to an ellipse (Sect.~\ref{mapping}).}
\end{center}
\end{figure}

We found that this function is weakly dependent on $R_\lambda$. For example,  the absolute difference
$|\varepsilon_{\rm max}(\kappa_s^\varphi, R_\lambda=\infty) - \varepsilon_{\rm max}(\kappa_s^\varphi, R_\lambda=\pm R_{\rm th})|$ is at most 0.03 for $R_{\rm th}=4$, corresponding to a fractional difference of about 10\%, in the whole $\kappa_s^\varphi$ range. 
This maximum difference decreases to $0.01$ for $R_{\rm th} = 10$. Thus, 
it is sufficient  to define $\varepsilon_{\rm max}(\kappa_s^\varphi):=  \varepsilon_{\rm max}(\kappa_s^\varphi, R_\lambda=\infty)$ in the arc formation region. This function is shown in Fig.~\ref{gof2}. Clearly 
$\varepsilon_{\rm max}$ depends on $\kappa_s^\varphi$, decreasing for high characteristic convergences. For low values of $\kappa_s^\varphi$, $\varepsilon_{\rm max}$ converges to $0.5$. In Appendix~\ref{useful-fit1} we provide a best-fitting function to $\varepsilon_{\rm max}(\kappa_s^\varphi)$, which could be useful, for example, to check if a given solution from inverse modeling has a physically meaningful mass distribution. The limits on the shape of the iso-convergence contours will be revisited in the next section by measuring the deviation with respect to an elliptical shape.

\section{\label{mapping} Ellipticity of the PNFW mass distribution}
\begin{figure}[!htb]
 \begin{center}
 \sidecaption
 \resizebox{\hsize}{!}{\includegraphics{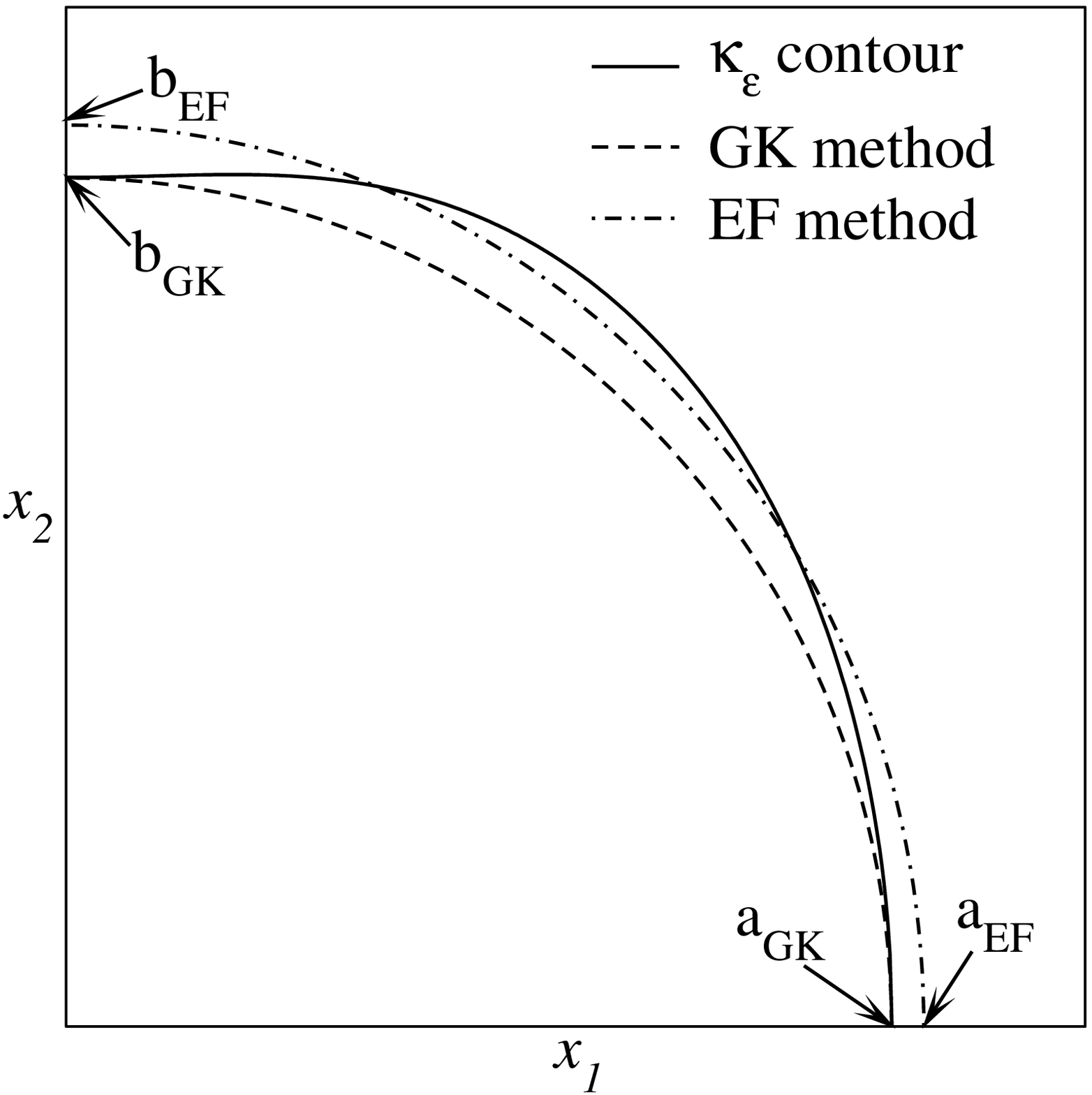}} \caption{ Illustration of the methods used to associate the PNFW mass  distribution to elliptical contours.  In the GK method, $a_{\rm GK}$ and $b_{\rm GK}$ correspond to the semi-major and semi-minor axes of the ellipse (dashed line) passing through the intersection of the iso-convergence contour (solid line) with the $x_1$ and $x_2$ axes. In the EF method, $a_{\rm EF}$ and $b_{\rm EF}$  correspond to the semi-major and semi-minor axes of the best-fitting ellipse (dash-dotted line) obtained by minimizing Eq. (\ref{gof-ef}). } \label{scheme}
 \end{center}
 \end{figure}

We adopted two procedures to associate an ellipticity $\varepsilon_\Sigma$ to each $\kappa_\varepsilon$ contour and to measure its deviation from an elliptical shape. In the first, we followed GK02 and define the semi-major axis $a_{\rm GK}$ and semi-minor axis $b_{\rm GK}$ as the intersections of the $\kappa_\varepsilon$ contour with the $x_1$ and $x_2$ axes, respectively (see Fig.~\ref{scheme}), such that the ellipticity is
\begin{equation}
 \varepsilon_\Sigma^{\rm GK}:=1-\frac{b_{\rm GK}}%
 {a_{\rm GK}}.
 \label{elip_gk}
 \end{equation}
 
The second procedure is to fit the $\kappa_\varepsilon$ contour by an ellipse. 
For this sake we introduce a figure-of-merit that represents the mean weighted squared fractional radial difference between the contour and the ellipse
\begin{equation}
\mathcal{D}^ 2  :=\frac{\sum_{i=1}^N\,w_i[r(\phi_{i})-r_\Sigma(\phi_{i}) ]^2 }%
  {\sum_{i=1}^N w_i\, r^2(\phi_{i})},
  \label{gof-ef}
 \end{equation}
where $N$ is the number of points on the $\kappa_\varepsilon$ contour, $\phi_i$ is their polar angle,
$w_i=\phi_i-\phi_{i-1}$ is a weight accounting for a possible non-uniform distribution of 
$\phi_i$, $r(\phi_{i})$ is the radial  coordinate of the $\kappa_\varepsilon$ contour and $r_\Sigma(\phi_{i})$ is  the radial coordinate of the ellipse, given by
 \begin{equation}
   r_\Sigma=\left[\left(\frac{\cos{\phi}}{a}\right)^2+%
   \left(\frac{\sin{\phi}}{b}\right)^2\right]^{-1/2},
   \label{first-elipse}
  \end{equation}
where $a$ and $b$ are the semi major and minor axes, respectively. The form of $\mathcal{D}^2$ in Eq.~(\ref{gof-ef}) was chosen to be scale-invariant and independent of the discretization\footnote{A convergence to within about $1\%$ is achieved for $N=100$ for the parameter ranges considered here.}. Owing to the symmetry of the $\kappa_\varepsilon$ contour, it is sufficient to compute $\mathcal{D}^2$ in the first quadrant of the lens plane.

The best-fitting ellipse is found by minimizing $\mathcal{D}^2$, for which we used the MINUIT code \citep{james92}. The resulting best-fitting values from this elliptical fit (EF) method, $a_{\rm EF}$ and $b_{\rm EF}$,  yield the ellipticity
 \begin{equation}
  \varepsilon_\Sigma^{\rm EF}:=1-\frac{b_{\rm EF}}%
 {a_{\rm  EF}}.
  \label{elip-ef}
 \end{equation}

When the iso-convergence contours are close to elliptical, we expect the results from both methods to be very similar. However, differences could emerge for high values of $\varepsilon$, especially in the dumbbell-shape regime. Another difference that will be discussed in Sect.~\ref{charac_convg_relation} arises when assigning a $\kappa_\varepsilon$ contour to the derived ellipse.

\begin{figure*}[!htb]
 \begin{center}
 \sidecaption
 \resizebox{\hsize}{!}{\includegraphics{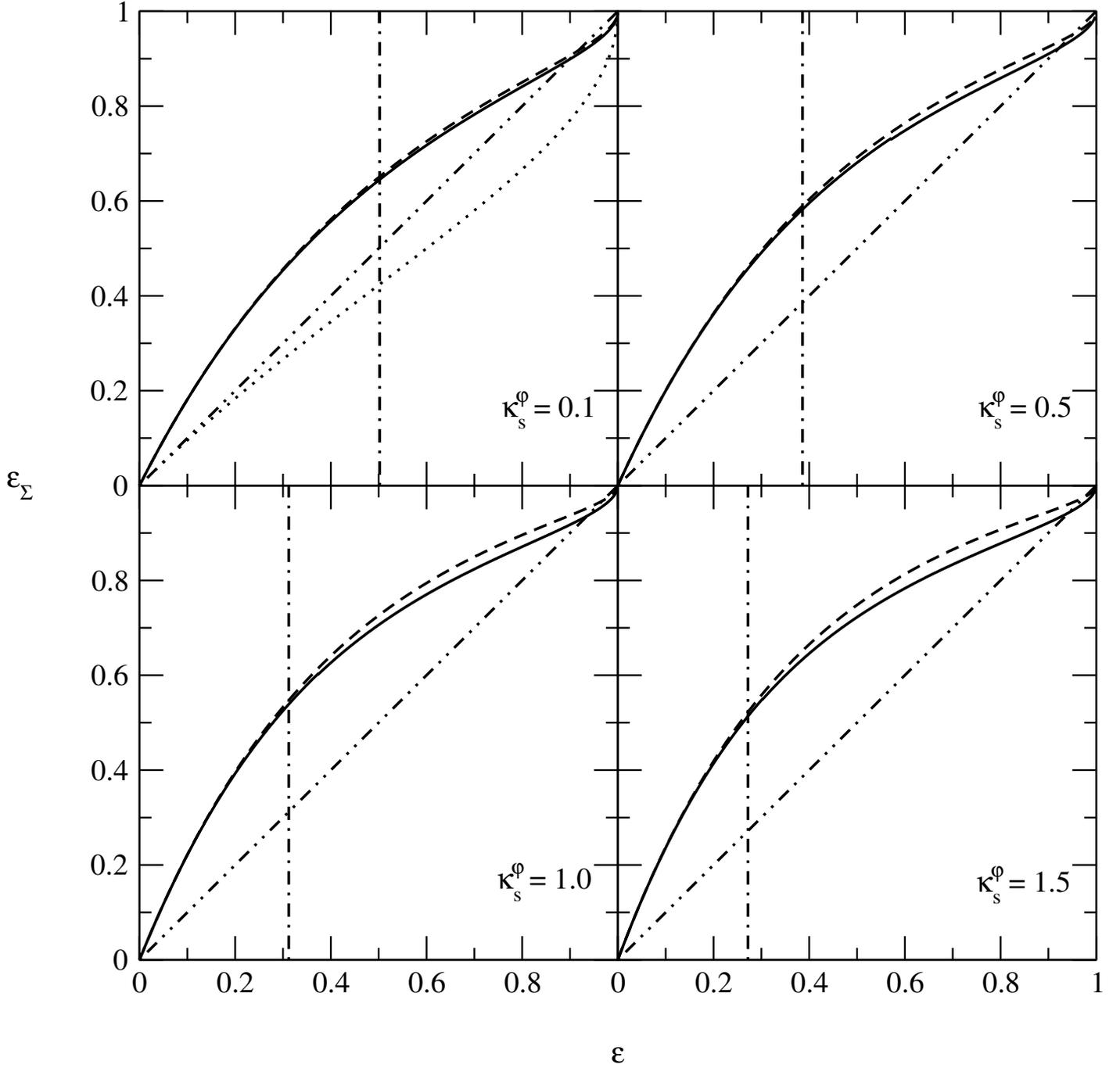}}\caption{\label{efmethod} Ellipticity of the PNFW mass distribution $\varepsilon_\Sigma$ from the elliptical fit ($\varepsilon_\Sigma^{\rm EF}$, solid lines) and GK ($\varepsilon_\Sigma^{\rm GK}$, dashed lines) methods, as a function of the parameter $\varepsilon$ for  four values of $\kappa_s^\varphi$, calculated at the tangential critical curve
 ($R_\lambda=\infty$) at $x_2=0$.  The dotted line  in the upper left panel corresponds to the ellipticity of the lensing potential $\varepsilon_\varphi(\varepsilon)$, Eq.~(\ref{ephi}). The dot-dot-dashed line shows the $\varepsilon_\Sigma=\varepsilon$ line to guide the eye.  The vertical dot-dashed lines show the values of $\varepsilon_{\rm max}(\kappa^\varphi_s)$.} 
 \end{center}
 \end{figure*}

We may associate an ellipticity to the iso-convergence contours related to each distortion curve. However, the functions $\varepsilon_\Sigma^{\rm GK,EF}(\varepsilon,\kappa_s^\varphi, R_\lambda)$  are weakly dependent on $R_\lambda$ close to the arc formation region. For example, the absolute differences
$\left|\varepsilon_\Sigma^{\rm  GK,EF}(\varepsilon,\kappa_s^\varphi,R_\lambda=\infty) -\varepsilon_\Sigma^{\rm GK,EF}(\varepsilon,\kappa_s^\varphi,\pm R_{\rm th})\right|$ 
are at most $0.01$ for $R_{\rm th}=10$ and $0.02$ for $R_{\rm th}=4$. Therefore, we chose the ellipticity associated to the convergence at the critical curve  ($R_\lambda=\infty$) as the ellipticity in the arc formation region. In Fig.~\ref{efmethod} the resulting values for  $\varepsilon_\Sigma^{\rm GK}$ (dashed lines) and $\varepsilon_\Sigma^{\rm EF}$ (solid lines) are shown as a function of $\varepsilon$, for
$\kappa_s^\varphi=0.1,0.5$, $1.0$, and $1.5$.  

As expected, the two ellipticity measures agree very well for low values of $\varepsilon$. For example, for  $\varepsilon = 0.25$,  $\left|\varepsilon_\Sigma^{\rm EF}-\varepsilon_\Sigma^{\rm GK}\right|$  is at most $0.03$ on the whole $\kappa_s^\varphi$ range. As can be seen in Fig.~\ref{efmethod}, although the behavior of the two functions is qualitatively very similar in the whole ellipticity range, noticeable differences appear precisely for the values of $\varepsilon$ (shown in Fig.~\ref{gof2}) close to where the dumbbell-shape arises.
 
The minimum value of the figure-of-merit (\ref{gof-ef}) $\mathcal{D}^2_{\min}$
can be used as a goodness of fit and hence as an estimator of the departure from the elliptical shape.
By setting a threshold on $\mathcal{D}^2_{\min}$ a maximum value of  $\varepsilon$ can be determined, for each $\kappa_s^\varphi$, such that $\mathcal{D}^2_{\min}$ does not exceed this threshold, ensuring a small deviation from the elliptical shape. 
In particular, setting this threshold at  $4.5 \times 10^{-4}$ avoids the dumbbell-shaped mass distribution, as shown in Fig~\ref{gof2}. Thus,  a small deviation from the elliptical shape is deeply connected to the avoidance of dumbbell shapes. Both conditions impose similar restrictions on the ellipticity and imply that the PNFW could be used to model the mass distribution in the region of arc formation for $\varepsilon$ below the limits given in Fig.~\ref{gof2} and Eq.~(\ref{emax_fit_funct}). 

The validity of the PNFW to represent elliptical mass distributions was also investigated in GK02. They considered a single value of the characteristic convergence and investigated the shape of the mass distribution as a function of the distance of the $\kappa_\varepsilon$ contour to the lens center. They provide a fitting function for $\varepsilon_{\Sigma}$ as a function of $x_{\rm GK}=\sqrt{a^2_{\rm GK}+b^2_{\rm GK}}$ and $\varepsilon$, valid for $\varepsilon < 0.25$.
Their lensing potential depth is expressed in terms of a characteristic velocity $v_c$,  connected to the NFW profile parameters by  \citep{golse2002}
\begin{equation} 
v_c^2 = \frac{8}{3}G\rho_s r_s^2.
\end{equation}
Using the values in GK02 ($v_c=2000\, {\rm km/s}$, $r_s=150\,  {\rm kpc}$, $z_L=0.3$, and $z_S=1$),  assuming $\Omega_m=0.3$, $\Omega_{\Lambda}=0.7$, and $H_0 = 65 \, {\rm km/s/Mpc}$, and using Eqs. (\ref{ks_nfw}) and (\ref{Sigmacrit}) yields $\kappa_s^{\varphi} \simeq 0.88$. Our results for $\varepsilon^{\rm GK}_{\Sigma}(\varepsilon, \,\kappa_s^{\varphi}=0.88)$ reproduce their fit to within 2\% down to its limit of validity.

On the other hand, by exploring a large interval of $\kappa_s^{\varphi}$, we find that higher values of 
$\varepsilon$ may be allowed, at least in the region of arc formation. In particular, values as high as $\varepsilon \simeq 0.5$ (corresponding to $\varepsilon_\Sigma \simeq 0.65$, see Fig.~\ref{efmethod}) are permitted for low  $\kappa_s^{\varphi}$. It is therefore useful to provide a fit that is valid beyond $\varepsilon = 0.25$ and includes the dependence of the ellipticity with $\kappa_s^{\varphi}$. Such a fitting function for  $\varepsilon^{\rm EF}_{\Sigma}(\varepsilon, \kappa_s^{\varphi})$ is shown in Appendix~\ref{useful-fit2}, which is valid in the whole range of $\varepsilon$ and the range of $\kappa_s^{\varphi}$ considered in this paper.

\section{\label{charac_convg_relation} Mapping among the PNFW and ENFW models} 

The ENFW model is constructed by replacing $\kappa(x)$ in Eq. \ref{kappa_nfw}
by $\kappa(x_{\varepsilon_{\Sigma}})$ (see Sect. \ref{pemodels}), where $x_{\varepsilon_{\Sigma}}$ is chosen as \citep{caminha09}
\begin{equation}
x^2_{\varepsilon_{\Sigma}} = \left(1-\varepsilon_{\Sigma}\right) x_1^2+  \left(\frac{1}{1-\varepsilon_{\Sigma}}\right) x_2^2
\end{equation}
such that $\varepsilon_{\Sigma}$ is the ellitpicity of the mass distribution. In this case we denote the NFW 
characteristic convergence by $\kappa_s^\Sigma$.

We constructed a mapping among the PNFW and ENFW models such that their mass distribution is similar on the arc formation region. In other words, for each pair ($\kappa_s^\varphi$, $\varepsilon$) we associated a corresponding pair of the ENFW model parameters ($\kappa_s^\Sigma$, $\varepsilon_\Sigma$), where $\kappa_s^\Sigma$ is the characteristic convergence of the associated ENFW model. 
The ellipticity of the ENFW mass distribution $\varepsilon_\Sigma$ is simply given by the ellipticity  associated to the $\kappa_\varepsilon$ contours, as discussed in Sect.~\ref{mapping}, with approximate expressions given in Appendix~\ref{useful-fit2}. 

The determination of $\kappa_s^\Sigma$  is numerically more complex and involves an ambiguity on how to perform the matching among models. We have chosen to perform the matching at  $a_{\rm EF}$, i. e., at the intersection of the best-fitting ellipse to the $\kappa_\varepsilon$ contour associated to a given $R_\lambda = const.$ curve with the $x_1$ axis. We have considered two possibilities: {\it i}) matching the value of the ENFW convergence at $a_{\rm EF}$ with the PNFW convergence associated to the  $R_\lambda = const.$ curve, 
{\it ii}) matching the position of $R_\lambda = const.$ curve of the ENFW model, i.e. such that it intersects the $x_1 $ axis at $a_{\rm EF}$. For any pair ($\kappa_s^\varphi$, $\varepsilon$) we fixed the ellipticity of the ENFW model  as $\varepsilon_\Sigma(\varepsilon,\kappa_s^{\varphi})$ and obtained $\kappa_s^\Sigma$ following the two procedures above.

The two possibilities described above yield very similar values for $\kappa_s^\Sigma$ down to high ellipticites. For example, taking as reference the $\kappa_\varepsilon$ associated to the critical curve (i.e. to $R_\lambda = \infty$), the maximum difference of $\kappa_s^\Sigma$ among the two procedures for  $\kappa_s^\varphi = 1.5$ and  $\varepsilon = 0.8$  is 3\% and this difference decreases substantially for lower ellipticities and $\kappa_s^\varphi $. On the other hand, if instead of using $a_{\rm EF}$ as reference position for the association of $\kappa_s^\Sigma$ we use $a_{\rm GK}$, the results are still very similar for $\varepsilon < \varepsilon_{\max} (\kappa_s^\varphi)$, but may differ substantially for higher ellipticities.

As described above, convergences can be matched for $\kappa_\varepsilon$ contours associated to any $R_\lambda$ curve.
We verified that the derived characteristic convergence is almost constant in the region of tangential arc formation. Indeed, the absolute differences $|\kappa_s^\Sigma(\varepsilon,\kappa_s^\varphi,R_\lambda=\infty)- \kappa_s^\Sigma(\varepsilon,\kappa_s^\varphi,R_\lambda=\pm R_{\rm th})|$  are at most $0.005$ ($0.01$), for $\kappa_s^\varphi \lesssim 0.05$ and $0.01$ ($0.03$), for $\kappa_s^\varphi \sim 1.5$, with $R_{\rm th} =10$ ($R_{\rm th}=4$), and  $\varepsilon < 0.6$. Therefore, as in the $\varepsilon_\Sigma$ case, we chose the characteristic convergence in the region of arc formation as the value of $\kappa_s^\Sigma(\varepsilon,\kappa_s^\varphi)$ calculated at the intersection of $R_\lambda = \infty$ with the $x_1$ axis.  As will be discussed in Sect. \ref{compar_sc}, we chose method ({\it ii}) to make the association between the two models.

\begin{figure}[!ht]
 \begin{center}
 \sidecaption
 \resizebox{\hsize}{!}{\includegraphics{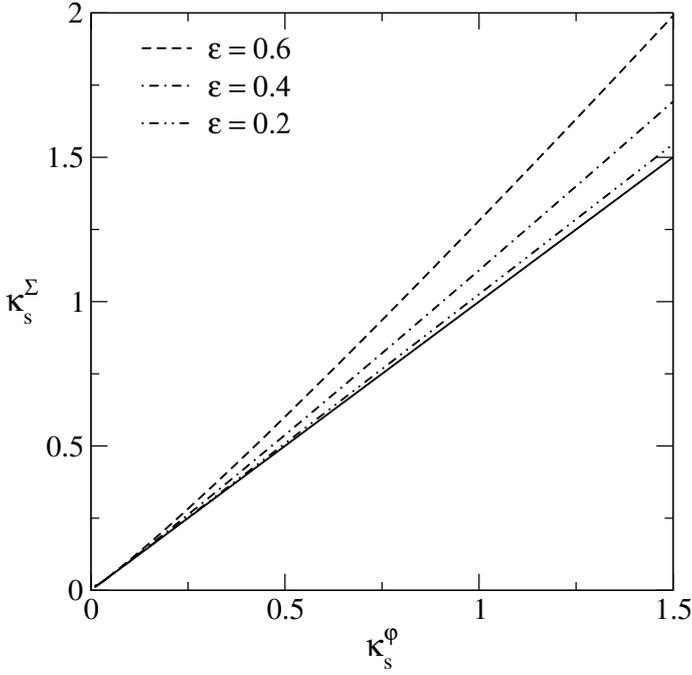}}
 \caption{\label{ks_relation_plot} Relation among the characteristic convergences of the PNFW and ENFW models  for some values of the parameter $\varepsilon$  calculated at the tangential critical curve ($R_\lambda=\infty$) at $x_2=0$. The solid line shows the $\kappa_s^\Sigma=\kappa_s^\varphi$ line to guide the eye. }
 \end{center}
 \end{figure}

In Fig.~\ref{ks_relation_plot} we show $\kappa_s^\Sigma$ as a function of $\kappa_s^\varphi$ 
for some values of $\varepsilon$. As expected, for $\varepsilon=0$ 
we have $\kappa_s^\Sigma=\kappa_s^\varphi$, but this equality does not hold for non-zero ellipticities. 
In particular, $\kappa_s^\Sigma$ is always larger than its corresponding $\kappa_s^\varphi$,  
and the difference between them increases with  $\varepsilon$. In Appendix~\ref{useful-fit3} we provide a fitting function for $\kappa_s^\Sigma(\varepsilon,\kappa_s^\varphi)$.

\section{\label{compar_sc}Comparison of the arc cross section}

The efficiency of a lens to produce arcs is quantified by the \textit{arc cross section} $\tilde{\sigma}_{R_{\rm th}}$, which is defined as the area in the source plane that generates images with $ L/W  \geq R_{\rm th}$, weighted by the multiplicity of the images \citep[i.e. multiply imaged regions are counted multiple times, see e.g.,][]{mene03}.
The computation of the arc cross section usually requires extensive arc simulations, which are computationally expensive \citep{miralda93b,bartelmann94,mene01,mene03,oguri03}.
However, in the infinitesimal circular source approximation, Eq.~(\ref{lw}), the cross section can be obtained directly from the local mapping from lens to source plane. In this case  $\tilde{\sigma}_{R_{\rm th}}$ is easily computed in the lens plane (in dimensionless coordinates) by  \citep[see, e.g.,][]{fedeli05,caminha09}
\begin{equation}
\tilde{\sigma}_{R_{\rm th}}=\int_{|R_\lambda|\geq R_{\rm th}}|\mu(\vec{x})|^{-1} d^2x,
\label{sigma}
\end{equation}
where $\mu=\left(\lambda_r \times \lambda_t\right)^{-1}$ is the magnification and  the integral is performed over a region with local distortion above the given threshold (i.e., the arc formation region introduced in Sect.~\ref{phys_lim_pnfw} and used in the preceding sections). 
Since the mapping between the PNFW and ENFW models was constructed in that region, the cross section 
is well suited to check if this mapping, obtained from matching the mass distribution, also holds for other lensing quantities, in this case the magnification. 

If the PNFW can be used to replace the ENFW in some applications (for example, arc statistics), we would expect the predictions for $\tilde{\sigma}_{R_{\rm th}}$ to be similar for both models, at least in the region where the PNFW provides an adequate description for the mass distribution. On the other hand, if the predictions do not match in this region, they can be used to set additional constraints on the PNFW model.

\begin{figure*}[!ht]
\centering \sidecaption \resizebox{\hsize}{!}{
\subfigure{\includegraphics{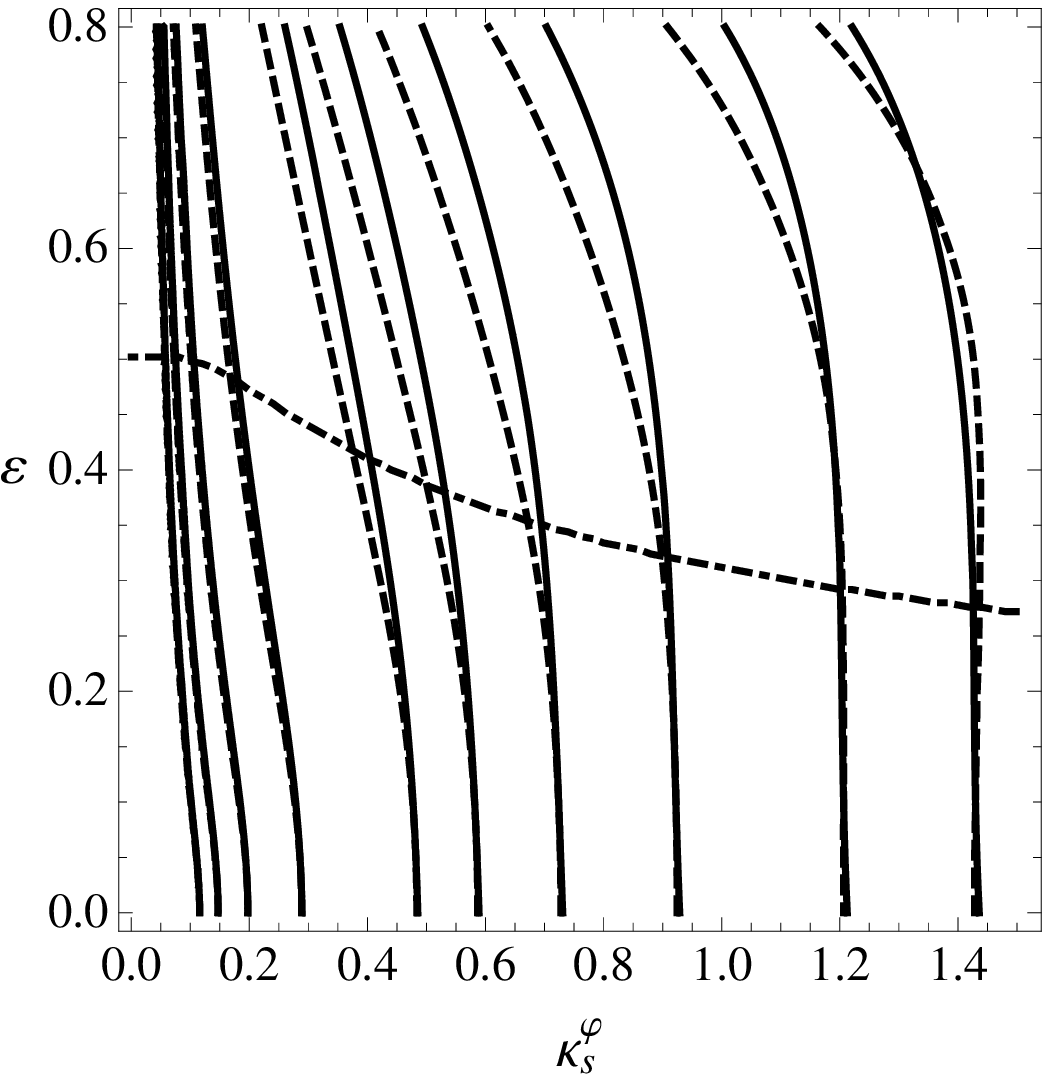}} \hspace{1.cm}
\subfigure{\includegraphics{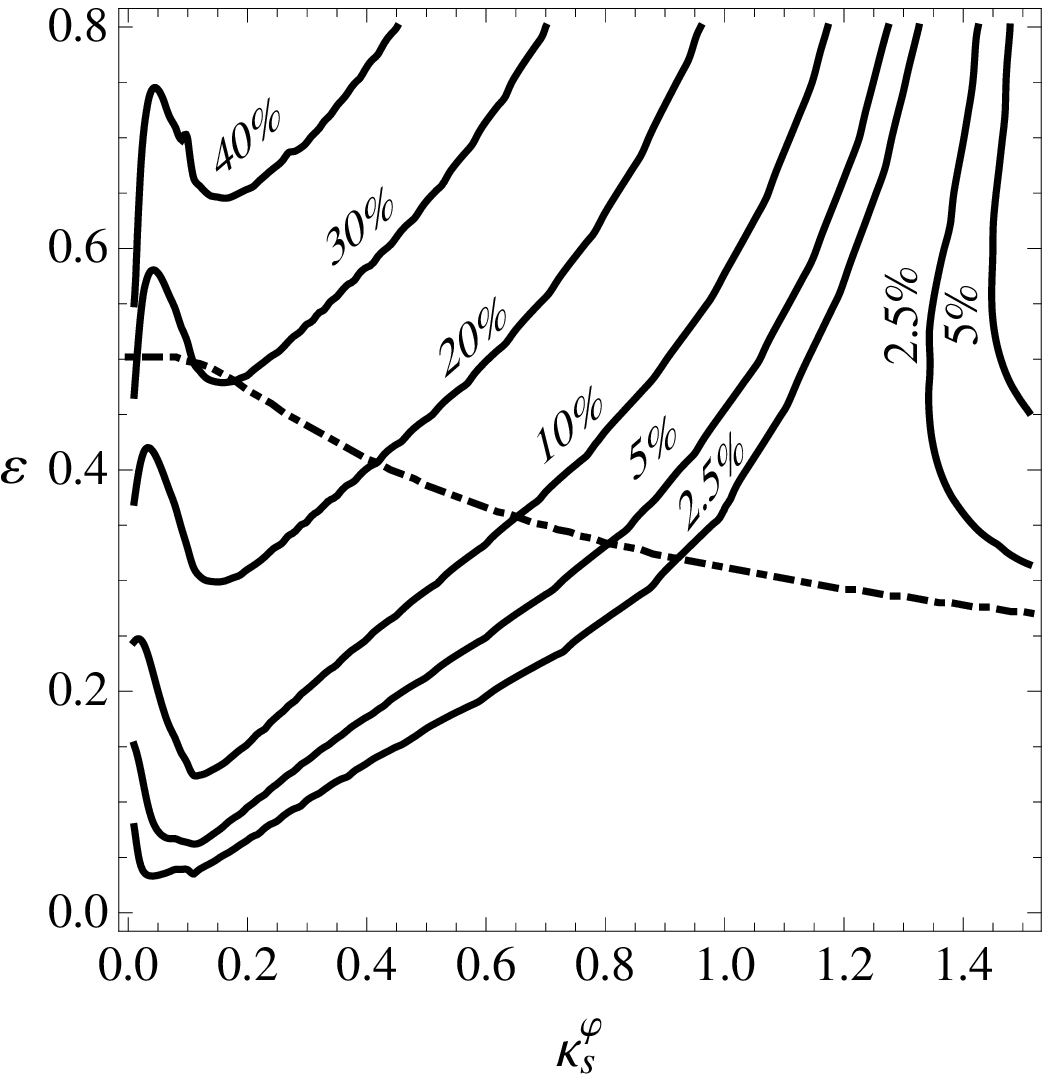}}}
\caption{\label{sigma_compar} Comparison between arc cross sections. Left panel: Contours of constant arc cross section in terms of the PNFW parameters. Solid lines correspond to $\tilde{\sigma}_{\rm ENFW}$ and dashed lines correspond to $\tilde{\sigma}_{\rm ENFW}$. 
The contours of constant arc cross section from left to right are $1 \times 10^{-6},1\times10^{-5}, 1\times 10^{-4}, 10^{-3}, 0.01,0.02,0.04,0.08,0.16$ and $0.24$. Right panel: Contours of  $\Delta\tilde{\sigma}/\tilde{\sigma}$ (Eq.~\ref{dif_rela_sigma}). In both plots, the dash-dotted line shows $\varepsilon_{\rm max} (\kappa_s^\varphi)$. Calculations were made for $R_{\rm th}=10$. }
\end{figure*}

To compare the cross section for both models we defined a regular grid in the ($\kappa_s^\varphi, \varepsilon$)
parameter space and mapped each point to ($\kappa_s^\Sigma,\varepsilon_\Sigma$) using the procedures described in section \ref{charac_convg_relation}. The cross sections $\tilde{\sigma}_{\rm PNFW}$ and $\tilde{\sigma}_{\rm ENFW}$ were computed for each set of parameters from Eq.~(\ref{sigma}) using the expressions for $\lambda_r(\vec{x})$ and $\lambda_t(\vec{x})$ of the corresponding model. The computation for the ENFW model was taken from \citet{caminha09}. The result for both cross sections is shown in the left panel of Fig.~\ref{sigma_compar} where contours of constant $\tilde{\sigma}_{R_{\rm th}}$ are displayed.
Visually, the arc cross sections of both models are similar in the region $\varepsilon<\varepsilon_{\rm max}$.

To quantify the difference between $\tilde{\sigma}_{\rm ENFW}$ and $\tilde{\sigma}_{\rm PNFW}$, we computed the relative difference
\begin{equation}
\frac{\Delta \tilde{\sigma}}{\tilde{\sigma}}=\left|\frac{\tilde{\sigma}_{\rm ENFW}-\tilde{\sigma}_{\rm PNFW}}{\tilde{\sigma}_{\rm ENFW}} \right|\label{dif_rela_sigma}.
\end{equation}

We have computed this fractional difference using the various possibilities for associating ENFW model parameters to PNFW parameters discussed in section \ref{charac_convg_relation}. Although those definitions generally yield similar values for $\kappa_s^\Sigma$, the value of $\Delta \tilde{\sigma}/\tilde{\sigma}$ can vary substantially, since the cross section is very sensitive to $\kappa_s$. Comparing the matches at $a_{\rm EF}$ and $a_{\rm GK}$, the latter gives a fractional difference at least 50\% higher than the former and this difference increases substantially with  $\kappa_s^\varphi$. The difference between the methods ({\it i}) and ({\it ii}) is much smaller and is not very sensitive to $\kappa_s^\varphi$, but in general procedure ({\it ii}) leads to smaller differences in the cross section than method ({\it i}). We chose therefore to define the matching among models using method ({\it ii}) at $a_{\rm EF}$ and compared the cross sections using this choice.

In the right panel of Fig.~\ref{sigma_compar} the contours of constant relative difference are shown. Our results show that for $\varepsilon<\varepsilon_{\rm max}$, $\Delta\tilde{\sigma}/\tilde{\sigma}$ can be as high as $30\%$ for low values of $\kappa_s^\varphi$. Thus, even in the region where the ENFW and PNFW mass distributions are similar, there can be substantial deviations on the cross section.

We may combine the constraints from the shape of the mass distribution and by assuming a maximum fractional deviation for the cross section. For example, for a maximum deviation of $10\%$, the joint constraint leads to a region limited approximately by the lines
\begin{equation} \label{reg_lt_10}
\varepsilon = \left\{\begin{array}{lc} 
0.25-1.30\kappa^\varphi_s, & \kappa_s^\varphi <  0.1\\
0.08+ 0.42\kappa^\varphi_s, & 0.1 \leq \kappa_s^\varphi \leq  0.65, \\ 
0.41-0.09\kappa^\varphi_s, & \kappa_s^\varphi > 0.65. \end{array} \right.
\end{equation}
Within this region the PNFW model can reproduce both the local mapping and the mass distribution of the ENFW model.

\section{\label{s&c}Summary and concluding remarks}

Motivated by its potential applications for gravitational arcs, we revisited the PNFW model, seeking to determine domains of validity in terms of the mass distribution and the arc cross section.
We have shown that the lensing functions of pseudo-elliptical models have simple analytic expressions (Eqs.~\ref{kappa_pnfw}--\ref{gamma_pnfw}) for any choice of the parameterization of the ellipticity.

We analyzed the PNFW mass distribution, in the ``arc formation region'' limited by the constant distortion curves ($R_\lambda=\pm R_{\rm th}$) by associating $\kappa_\varepsilon$ contours to these curves.
We verified that the convergence never takes negative values in this region. 
Since the results obtained in this work are weakly dependent on $R_{\rm th}$, for $R_{\rm th} > 4$, we  chose the critical curve ($R_\lambda \rightarrow \infty$) to derive the final relations summarized below.

We determined the maximum value of the potential ellipticity parameter, $\varepsilon_{\rm max}(\kappa_s^\varphi)$, such as to avoid dumbbell-shaped mass distributions. The results (Fig.~\ref{gof2}) enlarge the domain of applicability of the PNFW model to describe the mass distribution, at least in the arc formation region, beyond the commonly adopted value of $\varepsilon \simeq 0.25$, allowing values as high as $\varepsilon = 0.5$ for low values of $\kappa_s^\varphi$.

We introduced the figure-of-merit $\mathcal{D}^ 2$ (Eq.~\ref{gof-ef}) to quantify the deviation of the $\kappa_\varepsilon$ contours from the elliptical shape. Setting a maximum value for $\mathcal{D}^ 2$  at $4.5 \times 10^{-4}$ also avoids dumbbell-shaped mass distribution (Fig~\ref{gof2}), showing that the contours deviate from the elliptical shape on the verge of the emergence of the dumbbell shape.

The function $\mathcal{D}^ 2$ can be used to assign a best-fitting ellipse to the $\kappa_\varepsilon$ contour and hence to obtain an ellipticity $\varepsilon_\Sigma$ for the mass distribution (EF method). We showed that the  ellipticites $\varepsilon_\Sigma$ obtained from this method are almost identical to the method in GK02, especially for low ellipticities. However, the EF is better suited to assign an iso-convergence contour to the $\kappa$ contour. Furthermore, using the EF allows one to match between the PNFW and the ENFW models in a way that minimizes the difference between cross sections.

We provided fitting functions for $\varepsilon_\Sigma^{\rm EF}(\varepsilon,\kappa_s^\varphi)$ (Eqs.~\ref{analytical-expr}--\ref{ci_fit_funct}), extending the results of GK02, in the arc formation region, for any value of $\varepsilon$ and including the dependence on $\kappa_s^\varphi$. 
Going to higher ellipticities is relevant, given that values as high as $\varepsilon \simeq 0.5$ (corresponding to $\varepsilon_\Sigma \simeq 0.65$, see Fig.~\ref{efmethod}) are allowed in the arc formation region. 

From N-body simulations, it is found that the probability distribution of  the projected ellipticity 
peaks at about $\varepsilon_\Sigma=0.5$, with only a small fraction of the halos having $\varepsilon_\Sigma > 0.6$ \citep{oguri03}. Converting $\varepsilon_{\max}(\kappa_s^\varphi)$ to $\varepsilon_\Sigma$ we found that values of $\varepsilon_\Sigma > 0.5$ are allowed in the whole investigated range of $\kappa_s^\varphi$.
Therefore, the PNFW would provide a good description of the ENFW mass distribution for most expected values of  $\varepsilon_\Sigma$. 

By associating the iso-convergence contours of the PNFW to the ENFW close to the tangential critical curve, we obtained a relation among characteristic convergences $\kappa_s^\Sigma(\varepsilon,\kappa_s^\varphi)$. Fitting functions for this relation are provided in Appendix~\ref{useful-fit3} (for the case ({\it ii}) discussed in Sect. \ref{charac_convg_relation}). Combined with the relation among ellipticities, this function completes the mapping of
the parameters ($\varepsilon,\kappa_s^\varphi$) of the PNFW model to the parameters ($\varepsilon_\Sigma,\kappa_s^\Sigma$) of the ENFW model.

To test the mapping in a practical application we compared the predictions for a quantity that is useful in arc statistics. We computed the arc cross sections for both the PNFW and ENFW models and compared their predictions by matching the model parameters using this mapping. We did not find a direct connection between $\Delta\tilde{\sigma}/\tilde{\sigma}$ and $\varepsilon_{\rm max}$ (which has a similar shape as a function of $\kappa_s^\varphi$ as the  contours of constant  $\mathcal{D}^ 2$), although the cross section was computed in the region where the mapping is obtained. In other words, the limits derived from the shape of the mass distribution do not match those from cross section.

We may use this result to set additional constraints on the parameters of the PNFW model, by requiring 
an agreement with the ENFW for $\tilde{\sigma}$ in addition to the condition  $\varepsilon<\varepsilon_{\rm max}(\kappa_s^\varphi)$. This would ensure that the mass distribution as well as the local mapping (represented by the magnification $\mu$) are well reproduced by the PNFW.
Approximate limits of this combined restriction, imposing an agreement of about 10\% for the cross sections, are given in Eq.~(\ref{reg_lt_10}).

This new restriction should now be tested in other applications, especially with simulations using finite sources, to check if the PNFW and ENFW can be mapped to reproduce the same physical results.
This would validate the use of pseudo-elliptical models for simulations and the inverse problem, providing the relation to the associated elliptical model.

\begin{acknowledgements}
We thank the anonymous referee for useful comments and suggestions that led to considerable improvements on this manuscript.
H.~S.~D\'umet-Montoya is funded by CNPq  (PDJ/162989/2011-3), FAPERJ (``Nota 10'' fellowship, E-26/101.784/2010),  and the PCI/MCTI  program at CBPF (301.860/2011-4).  G.~B.~Caminha is funded by CNPq and CAPES. M.~Makler is partially supported by CNPq (grants 312876/2009-2 and 486138/2007-0) and FAPERJ (grant E-26/110.516/2012). We also acknowledge the support of the Laborat\'orio Interinstitucional de e-Astronomia (LIneA) operated jointly by the Centro Brasileiro de Pesquisas F\'isicas (CBPF), the Laborat\'orio Nacional de Computa\c c\~ao Cient\'ifica (LNCC) and the Observat\'orio Nacional (ON) and funded by the Ministry of Science, Technology and Innovation (MCTI). 
\end{acknowledgements}

\begin{appendix}

\section{Lensing functions for pseudo-elliptical models \label{pe_functions}}\label{ap1}
To derive the lensing functions for pseudo-elliptic models, it is useful to introduce the following coordinate transformation:
\begin{equation}
 x_1=\frac{x_\varepsilon}{\sqrt{a_{1}}}\cos{\phi_\varepsilon},\qquad
x_2=\frac{x_\varepsilon}{\sqrt{a_{2}}}\sin{\phi_\varepsilon},
\label{vx1-vxe}
\end{equation}
where $x_\varepsilon=\sqrt{a_{1}\,x_1^2+a_{2}\,x_2^2}$ and $\phi_\varepsilon=\arctan{(x_2/x_1\sqrt{a_{2}/a_{1}})}$. The Jacobian matrix of this transformation is
\begin{eqnarray}
\mathcal{J}(\vec{x},\vec{x}_\varepsilon)&=&\left[\!\begin{array}{c c} \frac{1}{\sqrt{{a_{1}}}}\cos{\phi_\varepsilon} & -\frac{x_\varepsilon}{\sqrt{{a_{1}}}}\sin{\phi_\varepsilon} \\
\frac{1}{\sqrt{{a_{2}}}}\sin{\phi_\varepsilon} & \frac{x_\varepsilon}{\sqrt{{a_{2}}}}\cos{\phi_\varepsilon} \end{array}\!\right].%
\label{jacob_transf-vx1-vxe}
\end{eqnarray}

Since the gradient operator transforms as $\nabla_{\vec{x}}=\mathcal{J}^{-1}(\vec{x},\vec{x}_\varepsilon)\,\nabla_{\vec{x}_\varepsilon}$, 
we have
\begin{eqnarray}
\partial_{x_1}&=&\sqrt{a_{1}}\cos{\phi_\varepsilon}\partial_{x_{\varepsilon}}-%
\frac{\sqrt{a_{1}}}{x_\varepsilon}\sin{\phi_\varepsilon}\partial_{\phi_\varepsilon}, \label{d1_e}\\
\partial_{x_2}&=&\sqrt{a_{2}}\sin{\phi_\varepsilon}\partial_{x_{\varepsilon}}+%
\frac{\sqrt{a_{2}}}{x_\varepsilon}\cos{\phi_\varepsilon}\partial_{\phi_\varepsilon}. \label{d2_e}
\end{eqnarray}

Using the expressions above, the deflection angle ($\vec{\alpha}_\varepsilon(\vec{x})=\nabla_{\vec{x}}\varphi(x_\varepsilon)$) for an elliptical potential reads (GK02)
\begin{equation}
\vec{\alpha}_\varepsilon(\vec{x})=  \left(\begin{array}{c} \alpha_{1}(\vec{x}) \\ \alpha_{2}(\vec{x}) \end{array}\right)=  \left(\begin{array}{c}
\alpha(x_\varepsilon)\sqrt{a_{1}}\cos{\phi_\varepsilon}\\
\alpha(x_\varepsilon)\sqrt{a_{2}}\sin{\phi_\varepsilon}
\end{array}\right), \label{ang_pel2}
\end{equation}
where $\alpha(x_\varepsilon)$ is the deflection angle of a circular model evaluated at $x=x_\varepsilon$.

Taking the partial derivatives of each component of the angle deflection and using Eqs.~(\ref{d1_e}~--~\ref{d2_e}) it
is straightforward to obtain 
\begin{eqnarray}
\partial_{x_1}\alpha_{1}(\vec{x})&=& %
a_{1}\left[\frac{d\alpha(x_\varepsilon)}{d x_\varepsilon}\cos^2{\phi_\varepsilon}%
+\frac{\alpha(x_\varepsilon)}{x_\varepsilon}\sin^2{\phi_\varepsilon}\right], \label{comp_x_elad}\\
\partial_{x_2}\alpha_{2}(\vec{x})&=& %
a_{2}\left[\frac{d\alpha(x_\varepsilon)}{d x_\varepsilon}\sin^2{\phi_\varepsilon}%
+\frac{\alpha(x_\varepsilon)}{x_\varepsilon}\cos^2{\phi_\varepsilon} \right],\label{comp_y_elad}\\
\partial_{x_2}\alpha_{1}(\vec{x})&=&\frac{\sqrt{a_{1}a_{2}}}{2}%
\left[\frac{d\alpha(x_\varepsilon)}{d x_\varepsilon}-\frac{\alpha(x_\varepsilon)}{x_\varepsilon}\right]\sin{2\phi_\varepsilon}\nonumber \\
  &=& \partial_{x_1}\alpha_{2}(\vec{x}) \label{comp_xy_elad}.
\end{eqnarray}
Using the relations for the lensing functions for circular potentials 
\begin{eqnarray}
\kappa(x)=\frac{1}{2}\left[\frac{\alpha(x)}{x} + \frac{d\alpha(x)}{d x}\right],
\,
\gamma(x)=\frac{1}{2}\left[\frac{\alpha(x)}{x}- \frac{d\alpha(x)}{d x}\right],\label{gamma_circ}
\end{eqnarray}
it is possible to express Eqs.~(\ref{comp_x_elad})~--~(\ref{comp_xy_elad}) as a function of $\kappa(x_\varepsilon)$ and $\gamma(x_\varepsilon)$, i.e.
\begin{eqnarray}
\partial_{x_1}\alpha_{1}(\vec{x})&=& %
a_{1}\left[\kappa(x_\varepsilon)-\gamma(x_\varepsilon)\cos{2\phi_\varepsilon}\right], \label{comp_x_elad2}\\
\partial_{x_2}\alpha_{2}(\vec{x})&=& %
a_{2}\left[\kappa(x_\varepsilon)+\gamma(x_\varepsilon)\cos{2\phi_\varepsilon}\right],\label{comp_y_elad2}\\
\partial_{x_2}\alpha_{1}(\vec{x})&=&-\sqrt{a_{1}a_{2}}\gamma(x_\varepsilon)%
\sin{2\phi_\varepsilon}=\partial_{x_1}\alpha_{2}(\vec{x}). \label{comp_xy_elad2}
\end{eqnarray}
Applying the usual definitions for the convergence and shear in terms of the deflection angle, we obtain Eqs. (\ref{kappa_pnfw}-\ref{gamma_pnfw}).

\section{Fitting formulae for $\varepsilon_{\rm max}$ and for the PNFW---ENFW mapping\label{useful-fit}}

\subsection{\label{useful-fit1} Fitting formula for $\varepsilon_{\rm max}$.}

Applying the procedure outlined in Sect.~\ref{phys_lim_pnfw},  we obtained the maximum value $\varepsilon_{\rm max}$  to avoid the dumbbell-shaped mass distribution as a function of $\kappa_s^\varphi$. 
For $\kappa_s^\varphi < 0.1$ we have  $\varepsilon_{\rm max} = 0.5$ (corresponding to the plateau in Fig.~\ref{gof2}). For higher values of $\kappa_s^\varphi$ this function is well fitted by a Pad\'e approximant of the form

\begin{equation}
\varepsilon_{\rm max}(\kappa_s^\varphi)%
=\frac{\sum_{n=0}^{4}a_n(\kappa_s^\varphi)^n}{\sum_{m=0}^{2}b_m(\kappa_s^\varphi)^m},
\label{emax_fit_funct}
\end{equation}
where
$a_0= 0.502$, $a_1=-0.301$, $a_2= 0.043$, $a_3= 0.078$, $a_4=-0.037$ and  $b_0= 0.932$, $b_1= 0.092$, $b_2=-0.107$, which provides an excellent fit ($\chi^2 < 4 \times 10^{-6}$), for values of $\kappa_s^\varphi$ in the range $[0.1,1.5]$.
\subsection{Fitting formulae for the ellipticity of the mass distribution of the PNFW model }
\label{useful-fit2}

We have verified that the ratio $\varepsilon_\Sigma/\varepsilon$ is well-fitted by a third-order polynomial in $\varepsilon$ for the whole range of this parameter, 
\begin{equation}
 \varepsilon_\Sigma(\varepsilon,\kappa_s^\varphi) =
c_0(\kappa_s^\varphi)\varepsilon +c_1(\kappa_s^\varphi)\varepsilon^2%
+c_2(\kappa_s^\varphi)\varepsilon^3 +c_3(\kappa_s^\varphi)\varepsilon^4.
 \label{analytical-expr}
\end{equation}
The values of the coefficients $c_i$ are obtained from this fit for each $\kappa_s^\varphi$ in the considered range. These functions are in turn fitted by Pad\'e approximants of
the form
\begin{equation}
c_i(\kappa_s^\varphi)=\frac{\sum_{n=0}^{N}d_n(\kappa_s^\varphi)^n}{\sum_{m=0}^{M} e_m(\kappa_s^\varphi)^m },
\label{ci_fit_funct}
\end{equation}
where the coefficients $d_n$ and $e_m$ are given in Table~\ref{tab-c_i}.

\begin{table}[!ht]
\centering
\begin{tabular}{||c|c|c|c|c||}
\hline
\hline
       &          &                         &     &   \\
       & $c_0(\kappa_s^\varphi)$  & $c_1(\kappa_s^\varphi)$ &
$c_2(\kappa_s^\varphi)$  & $c_3(\kappa_s^\varphi)$\\
\hline
  $d_0$&  $2.523$   & $-0.380$ &   $0.701$       & $-0.005$   \\
  $d_1$&  $0.978$   & $-0.191$ &   $-1.384$      & $-0.014$   \\
  $d_2$&  $2.503$   & $-2.439$ &   $3.333$       & $0.110 $   \\
  $d_3$&  $0.303$   & $-0.540$ &   $0.188$       & $-0.217$   \\
  $d_4$&  $0.778$   & $\cdots$ &   $-0.699$      & $\cdots$   \\
\hline
  $e_0$&  $1.281$   & $0.183$  &   $0.585$       & $0.033$    \\
  $e_1$&  $0.492$   & $0.329$  &   $-0.360$      & $-0.079$   \\
  $e_2$&  $0.208$   & $0.669$  &   $1.684 $      & $0.361$    \\
  $e_3$&  $0.859$   & $\cdots$ &   $-0.773$      & $\cdots$   \\
\hline
$\chi^2$ & $7.9\times 10^{-6}$& $2.8\times 10^{-6}$  & $1.4\times 10^{-5}$& $6.7\times 10^{-7}$ \\
\hline
\hline
\end{tabular}
\caption{\label{tab-c_i} Results from the regression analysis using the Pad\'e approximant for $c_i(\kappa_s^\varphi)$. Polynomials of different degrees are used for  $c_0/c_2$ and $c_1/c_3$.  The last row corresponds to the values of $\chi^2$ for each function $c_i(\kappa_s^\varphi)$.}
\end{table}

We verified that the combination of \ref{analytical-expr} and \ref{ci_fit_funct} is indeed a good approximation for the ellipticity of the convergence contours in the region associated to the curves $R_\lambda = \pm R_{\rm th}$. 
For example, the absolute difference $\left|\varepsilon_\Sigma(\varepsilon,\kappa_s^\varphi) - \varepsilon_\Sigma^{\rm EF}(\varepsilon,\kappa_s^\varphi,R_\lambda=\pm 10)\right|$ is at most $10^{-3}$ in the whole range of $\varepsilon$ and $\kappa_s^\varphi$.

\subsection{Fitting formulae for mapping characteristic convergences}
\label{useful-fit3}
The relation between the characteristic convergence of the PNFW model
($\kappa_s^\varphi$) and the characteristic convergence of the ENFW model ($\kappa_s^\Sigma$) is fitted by a quadratic function in $\kappa_s^\varphi$
\begin{equation}
 \kappa_s^\Sigma(\varepsilon,\kappa_s^\varphi) =
p_0(\varepsilon)+p_1(\varepsilon)\kappa_s^\varphi
+p_2(\varepsilon)(\kappa_s^\varphi)^2.
 \label{analytical-expr_ks}
\end{equation}
By construction $p_0(\varepsilon)\rightarrow 0$, $p_1(\varepsilon)\rightarrow 1$, and $p_2(\varepsilon)\rightarrow 0$ as $\varepsilon\rightarrow0$.

 We divide the regression analysis into two ranges of $\kappa_s^\varphi$.  For $\kappa_s^\varphi\leq 0.1$ we choose $p_0(\varepsilon)=0$ and found that $p_1(\varepsilon)$ and $p_2(\varepsilon)$ are well-fitted by
\begin{equation}
p_1(\varepsilon)=\sum_{n=0}^4 g_n\varepsilon^n, \ \ p_2(\varepsilon)=\frac{h_0+h_1\varepsilon+h_2\varepsilon^2}{k_0+k_1\varepsilon},
\end{equation}
 with coefficients $g_n$, $h_n$ and $k_n$ given in Table \ref{tab_ks_lt01} (which are valid for $\varepsilon \leq 0.8$). We check the accuracy of Eq. (\ref{analytical-expr_ks}) with $p_1(\varepsilon)$ and $p_2(\varepsilon)$ given above by computing the $\chi^2$ for $\kappa_s^\varphi<0.1$ and $\varepsilon \leq 0.6$, and we found that it is less than $3\times10^{-9}$ in this range of parameter values.
\begin{table}[!ht]
\centering
  \begin{tabular}{||c|c|c|c||}
\hline
\hline
       & $p_1(\varepsilon)$ &  & $p_2(\varepsilon)$   \\
\hline
  $g_0$& $ 1.00   $    & $h_0$ &  $0.001 $ \\ 	
  $g_1$& $ 0.005  $    & $h_1$ & $ 0.446 $ \\
  $g_2$& $ -0.035 $    & $h_2$ & $-0.039  $ \\
  $g_3$& $ 0.316  $    & $k_0$ & $ 0.905 $  \\
  $g_4$& $ -0.257 $    & $k_1$ &  $-0.845$   \\
  \hline
$\chi^2$ & $1.25\times 10^{-8}$& &$2.56\times 10^{-7}$ \\
\hline
\hline
\end{tabular}
\caption{\label{tab_ks_lt01}  Results from the regression analysis using a polynomial
form for $p_1(\varepsilon)$ and a Pad\'e approximant for $p_2(\varepsilon)$.  The last row corresponds to the values of $\chi^2$ for each function.}
\end{table}

For $0.1<\kappa_s^\varphi \leq 1.5$ the functions $p_0(\varepsilon)$ and $p_2(\varepsilon)$ are well fitted by polynomials 
\begin{equation}
 p_{0,2}(\varepsilon)=\sum_{n=1}^{5}q_n\varepsilon^n,
\label{p_02}
\end{equation}
with coefficients $q_n$ given in Table \ref{tab_ks}, whereas $p_1(\varepsilon)$ is fitted by
\begin{equation}
 p_1(\varepsilon)=\frac{1+s_1\varepsilon+s_2\varepsilon^2+s_3\varepsilon^3}{1+t_1\varepsilon+t_2\varepsilon^2} \label{p1}
\end{equation}
with $s_1=-0.353, s_2=-0.0270, s_3=-0.133$ and $t_1=-0.339, t_2=-0.473$, which gives $\chi^2=8.55\times 10^{-7}$ for values of $\varepsilon \leq 0.8$ in the considered  $\kappa_s^\varphi$ range.

\begin{table}[!ht]
\centering
  \begin{tabular}{||c|c|c||}
\hline
\hline
       & $p_0(\varepsilon)$  & $p_2(\varepsilon)$   \\
\hline
  $q_1$& $0.008 $    &   $-0.030$ \\ 	
  $q_2$& $-0.091$   &   $ 0.596 $ \\
  $q_3$& $0.102$    &   $-0.970 $ \\
  $q_4$& $-0.052$   &   $ 0.787 $  \\
  $q_5$& $-0.034$   &   $0.056$   \\
\hline
$\chi^2$ & $1.25\times 10^{-7}$& $1.\times 10^{-6}$ \\
\hline
\hline
\end{tabular}
\caption{\label{tab_ks} Results from the regression analysis using polynomial
forms for $p_0$ and $p_2$.  The last row corresponds to the values of $\chi^2$ for each function.}
\end{table}

 We have also checked the accuracy of Eq. (\ref{analytical-expr_ks}) with $p_0(\varepsilon)$, $p_1(\varepsilon)$, and $p_2(\varepsilon)$ given above for values of the convergence close to the critical curves. We computed $\left|\kappa^\Sigma_s(\varepsilon,\kappa_s^\varphi) - \kappa^\Sigma_s(\varepsilon,\kappa_s^\varphi,R_\lambda=\pm 10)\right|$ 
and found that it is at most $10^{-2}$ for $ \varepsilon \leq 0.6$.  The $\chi^2$ of the fit is less than $9 \times 10^{-5}$ in the same parameter range.

\end{appendix}
\end{document}